\documentclass{bmcart}

\usepackage{url}
\usepackage{amssymb}

\usepackage{acro}
\DeclareAcronym{nntf}{
  short=NNTF,
  long=non-negative tensor factorization,
}
\DeclareAcronym{hcq}{
  short=HCQ,
  long=hydroxychloroquine
}
\DeclareAcronym{ihu}{
  short=IHU,
  long=French Marseille University Hospital Institute
}

\DeclareAcronym{ngo}{
  short=NGO,
  long=Non-Governmental Organization
}

\DeclareAcronym{who}{
  short=WHO,
  long=World Health Organization
}

\DeclareAcronym{ijaa}{
  short=IJAA,
  long=International Journal of Antimicrobial Agents
}

\usepackage{amsthm,amsmath}
\RequirePackage{hyperref}
\usepackage[utf8]{inputenc} 
\usepackage{url}
\usepackage{csquotes}

\usepackage{graphicx}

\startlocaldefs
\endlocaldefs

\begin{document}

\begin{frontmatter}

\begin{fmbox}
\dochead{Research}


\title{Temporal and geographic analysis of the Hydroxychloroquine controversy in the French Twittosphere}


\author[
  addressref={aff1},                   
  corref={aff1},                       
  email={mauro.fccn@gmail.com}   
]{\inits{M.}\fnm{Mauro} \snm{Faccin}}
\author[
  addressref={aff3},
  email={emilien.schultz@sciencespo.fr}
]{\inits{E.}\fnm{Emilien} \snm{Schultz}}
\author[
  addressref={aff4},
  email={floriana.gargiulo@cnrs.fr}
  ]{\inits{F.}\fnm{Floriana} \snm{Gargiulo}}


\address[id=aff1]{
  \orgdiv{DIMEC},             
  \orgname{University of Bologna},          
  \city{Bologna},                              
  \cny{Italy}                                    
}
\address[id=aff3]{
  \orgdiv{Médialab},             
  \orgname{Sciences Po},          
  \city{Paris},                              
  \cny{France}                                    
}
\address[id=aff4]{
  \orgdiv{Gemass},             
  \orgname{CNRS},          
  \city{Paris},                              
  \cny{France}                                    
}



\end{fmbox}


\begin{abstractbox}

\begin{abstract} 
  At the beginning of the COVID-19 pandemic, the urge to find a cure triggered an international race to repurpose known drugs.
  Chloroquine, and next hydroxychloroquine, emerged quickly as a promising treatment.
  While later clinical studies demonstrated its inefficacy and possible dangerous side effects, the drug caused heated and politicized debates at an international scale, and social media appeared to play a crucial role in those controversies.
  Nevertheless, the situation was largely different between countries.
  While some of them rejected quickly this treatment as France, others relied on it for their national policies, as Brazil.
  There is a need to better understand how such international controversies unfold in different national context.
  To study the relation between the international controversy and its national dynamics, we analyze those debates on Hydroxychloroquine on the French-speaking part of Twitter, focusing on the relation between francophone European and African countries.
  The analysis of the geographic dimension of the debate revealed the information flow across countries through Twitter's retweet hypergraph.
  Tensor decomposition of hashtag use across time points out that debates are linked to the local political choices.
  We demonstrate that the controversial debates find their center in Europe, in particular in France, while francophone Africa has a lower participation to the debates, following their early adoption of the familiar Hydroxychloroquine and rejection of WHO recommendations.
\end{abstract}




\end{abstractbox}
%

\end{frontmatter}





\section*{Context}
The recognition between January and February 2020 of the international dimension of the COVID-19 epidemic started a race to identify efficient therapeutic solutions.
Among the many attempts of drug repurposing, the proposal to use chloroquine first and then \ac{hcq} received exceptional attention~\cite{Gould2021,Schwartz2022}.
Preliminary results raised high expectations and attracted massively the attention of the scientific community, governments and general public.
This raised to numerous public debates on individual choice to use or not this therapeutic solution, the liberty of physicians to prescribe it, and how scientific proofs should inform political choices.
Due to their scale and their intensity, those debates built one of the most recent public controversy on health~\cite{Hallin2013}: scientists and physicians, but also citizens and political representatives, argued for or against the use of this drug, fostering circulation of contradictory claims in a context of scientific uncertainty.
As for other health controversies~\cite{Gargiulo2020}, social media have played a central role in the circulation of both scientific information and political debates, and participated to transform the hope for a cure to an \emph{infodemics}~\cite{Tuccori2020} characterized by a strong polarization~\cite{Rughinis2020}, especially on Twitter~\cite{Mutlu2020}.

The in vitro efficacy of chloroquine was first mentioned in the early 2020 in a Chinese scientific report leading to the onset of numerous clinical trials around the world to test the drug efficacy in different combinations~\cite{Pearson2021}.
The media coverage in the public sphere was largely triggered by Professor Didier Raoult and his team at the research center \ac{ihu}, in particular by their publication on the ``\ac{ijaa}'', dated 20th March 2020.
Along with post on Youtube platform and Twitter, this publication gave visibility to the promises of \ac{hcq} effectiveness in the treatment of COVID-19 (in combination with Azithromycin)~\cite{Schultz2021a}.
Following this first claim, Twitter played a central role~\cite{Marcon2021} for the diffusion of the topic outside the French borders.
This is best shown by Elon Musk's tweet:
\blockquote[Elon Musk, March 17, 2020]{%
  Hydroxychloroquine probably better.
}
followed by the US President Donald Trump:
\blockquote[US President Donald Trump, March 21, 2020]{%
  ``HYDROXYCHOLOROQUINE \& AZITHROMYCIN, taken together, have a real chance to be one of the biggest game changers in the history of medicine.
  The FDA has moved mountains --- Thank you! Hopefully they will BOTH (H works better with A, International Journal of Antimicrobial Agents) be put in use IMMEDIATELY.\@ PEOPLE ARE DYING, MOVE FAST, and GOD BLESS EVERYONE!''.
}
increasing the visibility of Professor Didier Raoult and boosting its position in France~\cite{Origgi2022}.
This important media coverage in both legacy media and social media had strong consequences on the following political decisions.
On the wave of this paper the governments of several counties allowed the use of \ac{hcq} for COVID-19 treatments,
while many clinical trials were started to assess \ac{hcq} efficacy.

Due to the time needed to achieve randomized clinical trials and deliver robust evidence, the uncertainty on \ac{hcq} efficacy persisted for several months which fueled debates.
In several countries, as the US or France, the debates were boosted by the political polarization between those who saw this treatment as a solution for the pandemic and those who doubted that it was a silver bullet and promoted other strategies to manage the virus, as lock-down, social distancing and masks~\cite{Rughinis2020}.
This division and the subsequent uncertainty was reinforced by the publication of an article in the Lancet journal the 22nd of May 2020 that claimed of \ac{hcq} inefficacy and even toxicity.
Clinical trials all around the world were stopped and countries, as France, withdrew the authorization for the clinical use of \ac{hcq} for COVID-19 patients, leading to a re-intensification of the debate on the topic.
The data of this article was then denounced as a fraud and the paper was retracted two weeks after~\cite{Krause2022}.
In particular a large Twitter activity has derived by the use of the Pr. Raoult's follower of the hashtag \texttt{\#LANCETGATE}, accusing the government to have stopped the use of \ac{hcq} based on a retracted paper.
Besides, those debates assumed extremely harsh tones, especially on social media, where many scientists and doctors were threatened and attacked~\cite{Ektorp2020, Schultz2021a}.

The controversy on \ac{hcq} had a worldwide reach, and had consequences on health behaviours, political decision and scientific organization.
One of its specificity was to bring debates on the functioning of research and the conditions of medical evidence, particularly on clinical trials, into the public and political debate.
For this reason, several social science and public health studies focused on this topic~\cite{Berlivet2020, Schultz2022a, Smyrnaios2021, Dubois2021}.
Those works showed that national specificity, such as the way government handled the pandemic, or the audience of central actors may have consequences on how the debate unfolded.
For instance, it appeared that the type of drug that attract the media attention can differ regarding the country~\cite{Bharti2022}.
In France, \ac{hcq} was presented by doctors and the media as a popular African drug~\cite{Desclaux2022}.
An initial fierce media and political debate on the scientific controversy, notably due to the communication capacity of the \ac{ihu} and Pr. Didier Raoult, was followed by a gradual dilution.
With the advent of vaccination and the government's strategy of prioritizing this crisis management, the promotion of \ac{hcq} becomes an element of the mobilizations that oppose this management~\cite{Schultz2021a}.

Although evidence has accumulated on the ineffectiveness of these molecules in treating this disease and even on their negative consequences~\cite{Axfors2021, Gould2021}, the drug were still administered in countries where national leaders have supported its use~\cite{Taccone2022}, leading to less public debates.
For instance, in Brazil, \ac{hcq} is integrated into the government's response to the crisis, to the extent that observers have spoken of a ``\emph{\acl{hcq} alliance}''~\cite{Casaroes2021}.
Many countries in West Africa integrated it into care protocols~\cite{Osuagwu2021}, notably because of the ``familiarity'' with this anti-malarial drug~\cite{Mukankubito2021}.
The Cameroonian government, for instance, recommended the use of \ac{hcq} and made it the national standard care.
The Ministry of Health promoted the national industry by ensuring an investment plan in the Institute for Research and Study of Medicinal Plants.
The debates in the public space focused on the issue of counterfeit drugs and, in a second time, on the scandal of a misappropriation of public money by the Minister of Health and the Minister of Scientific Research and Innovation, with few references to the WHO recommendation.

The case of \ac{hcq} allows us to directly address the issue of the circulation of debates in the context of the management of the COVID-19 pandemic, and to explore the relation between an international controversy and specific national debates.
While existing studies targeted either specific countries, as France, Brazil or the US, or a specific period of the controversy, we will focus on French-speaking countries and in particular to their relationship with France itself~\cite{Atlani-Duault2016}, the country that triggered the initial debate.
Moreover, while in the US~\cite{Rughinis2020} and Europe debates have been deeply discussed, little attention have been given to Africa.
If some observers claimed that \ac{hcq} debates was not a \enquote{French obsession}~\cite{Lapostolle2021}, there is little evidence of the extension of those debates in French-speaking countries.
On previous epidemics outbreaks, such as Ebola, the issue was specific to African countries~\cite{Roy2020}, limiting the possibility to make comparisons.
The particularity of the history of a repurposed drug for malaria, endemic in Africa, and the post-colonial history shared between France and several West African countries directly raises the question of the extension reached by the debates.

Based on a computational social science analysis of debates on a social media, Twitter, we propose to explore the scale effects of the debates that took place on \ac{hcq} between countries, with a focus on francophone African countries.
In this paper we will first analyze the temporal dimension of the discussion on \ac{hcq} in the French speaking Twittosphere.
We will show that the debates follow precise topic waves, that reflect the scientific and political agenda of the \ac{hcq} debate.
Secondly we will analyze the geographic distribution of the debate, showing a general national-level recirculation of the COVID-19 information structure, but also a high localization of the \ac{hcq} debate in the Northern Hemisphere.

\section*{Methods}

\subsection*{Data collection}

We collect tweets and retweets from Twitter using its search and stream APIs, leveraging the Gazouilloire~\cite{gazu} python software.
We used three set of keywords to independently extract tweets related to three topics: COVID-19, vaccines and \ac{hcq}, and merged the three initial datasets, keeping only the tweets mentioning either COVID-19 related keywords or \ac{hcq} related keywords.
For the full set of keywords, see~\cite{DataCovVac, faccin2022vaccinecritics}.
In this way we collected approximately 85M tweets from 768k users~\cite{DataCovHCQ}.

\subsection*{Geographic analysis}

A subset of Twitter users includes in their online profiles geographic data.
For each of them we checked their \emph{user description}, \emph{user real name} and \emph{user location} fields for the presence of geographic related keywords.
In particular, we searched for:
\begin{itemize}
  \item UTF-8 characters corresponding to a country flag.
  \item a capital's name,
  \item demonyms of a country,
  \item a country name in different languages.
\end{itemize}

Twitter users fitting in more than one country are ignored unless one of the two countries was France (often, France related keywords may refer to the spoken language).
In this way we geotagged approximately 380k users.

To further extend the initial set of geotagged users we considered the dilation of this set under the retweet network.
For each untagged user we consider the fraction of retweets towards or from users with a given geographic tag.
If this fraction surpasses 90\%, the user acquires the same geographic tag.
After this extension, geotagged users reached the number of approximately 488k.

This let us build a subset of tweets posted or shared by a tagged user linked to a given country.

\subsection*{Information flow across countries}

To study the information flow within or between countries we leverage the hypergraph representation of the retweet topology.
In particular, for any tweet in the dataset, we add a forward hyperedge~\cite{Ausiello_2017} between the tweeting user (hyperedge tail) and the retweeting users (hyperedge head)~\cite{faccin2022vaccinecritics, faccin2022directedhypergraphs}.
To simulate the information flow on such topology, we consider a Markov-chain such as a random walk.
The walker selects a hyperedge incident to its position with even probability and then jump to any of the head users with even probability.
The evolution of the random walker describes the information flow across the Twitter user base, and we can write the probability of reaching any user $j$ from any other user $i$, $p(i \to j) $, and the probability distribution $p(i)$ of visiting a user $i$ at the equilibrium.

Considering the partition of user into countries, one can compute country wide quantities such as the probability of the information to flow through a country (visiting probability):
\begin{equation}
  p_C = \sum_{i\in C}p(i),
\end{equation}
and the probability that information from country $C$ reach country $D$:
\begin{equation}
  p_{CD} = \sum_{i\in C, j\in D} p(i)p(i\to j).
  \label{eq:country_prob}
\end{equation}
Given the relations above, we compute the trapping probability as the probability of a piece of information to stay trapped within one of the partitions (in our case the partition of users into countries).
\begin{equation}
  p_C^\circlearrowleft  = \frac{p_{CC}}{p_C} = \frac{\sum_{ij \in C} p(i)p(i \to j)}{ \sum_{i\in C} p(i) }
  \label{eq:trapping}
\end{equation}
for any country $C$.

\subsection*{Topic detection with \acl{nntf}}

To extract the main topics and the evolution of the discussion across time and countries we extract the temporal hashtag co-occurrence and the country hashtag co-occurrence tensors.
The former tensor $T$ is defined when $T_{ijk}$ represents the number of tweets and retweets that hold both hashtags $i$ and $j$ on day $k$:
Similarly, the country hashtag co-occurrence tensor $K$ is defined when $K_{ijk}$ represents the number of tweets and retweets that hold both hashtags $i$ and $j$ and the tweeting or retweeting user is from country $k$.

To extract the time evolving topics we use the \ac{nntf}~\cite{Shashua_2005_nntf, Cichocki_2009_nn_factorization, M_rup_2011_application_factorizations} as in~\cite{Gauvin_2014_nntf_communities}.
In \ac{nntf}, the given tensor $T$ is decomposed in three non-negative matrices $A, B$ and $C$ where each entry of the tensor is as close as possible to:
\begin{equation}
  T_{ijk} \sim \sum_{c=1}^{N_C} A_{ic}B_{jc}C_{kc}
\end{equation}
where $N_C$ is the number of components fixed by the user.
In a symmetric framework as in our case, the first two matrices coincide $A=B$ and represent the hashtag composition of each component.
The third matrix $C$ represents the activity of each component along the third dimension, being it time or country.
Since the temporal/country activity of the components do not depend on the hashtag composition, we compute the strength of each component as follows:
\begin{equation}
  S_{kc} = C_{kc}\sum_{i=1}^N A_{ic}
\end{equation}
Where $N$ is the number of hashtags.

In the following we fix $N_C=9$ using the elbow criterion, although the results are robust to the variation of this parameter.

\section*{Results}
\subsection*{The timeline of the \ac{hcq} debate}

\begin{figure}[!htb]
  \begin{center}
    \includegraphics[width=\linewidth]{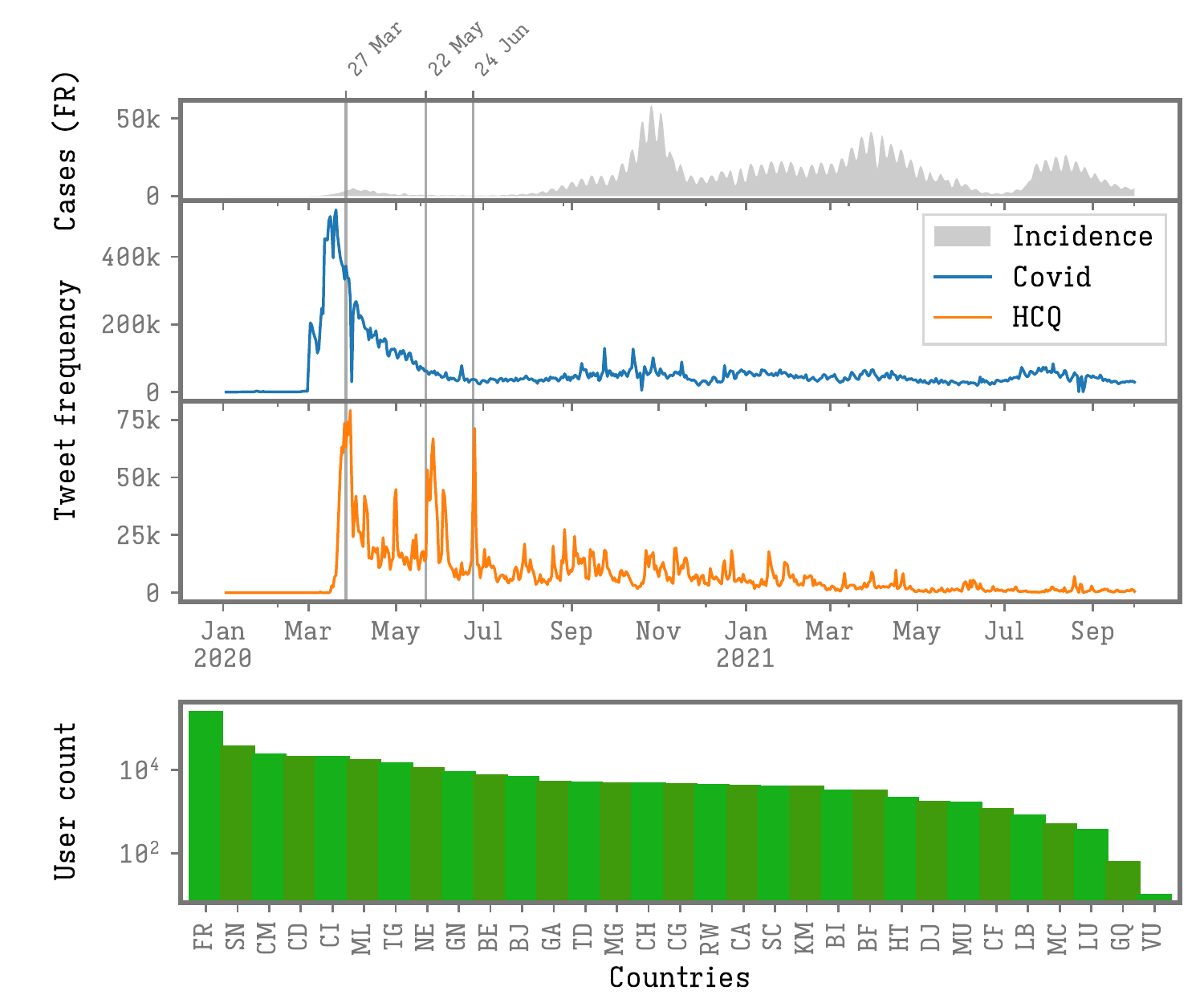}
  \end{center}
  \caption{%
    \csentence{Database statistics.}
    COVID-19 daily incidence in France for comparison (upper plot).
    Frequency of daily tweets and retweets on COVID-19 (second plot from above) or on \ac{hcq} (third plots).
    Number of users per country (lower plot).
  }\label{fig:stats}
\end{figure}

Fig~\ref{fig:stats} displays the tweeting frequencies for tweets containing either a keyword related to COVID-19 ($\sim$40M) or keyword related to \ac{hcq} ($\sim$5M).
While both timelines display an initial peak with a high volume of tweets, they also have specific characteristics.
The discussion related to  COVID-19 follows the epidemic peaks with a higher attention in the first pandemic period due to the emotional impact of the event.
The profile of the discussion on \ac{hcq} is influenced by the scientific and political agenda related to the debate.
During this period specific events influence the sudden user engagement in the \ac{hcq} discussion:
\begin{description}
  \item[March 27] The resonance of the publication of first Pr. Raoult's observational study (on March 20) and the political echo given by Donald Trump to this publication, generate a first attention peak in the debate;
  \item[May 22] Publication of the Mehra et al. article in the Lancet alerting about \ac{hcq} toxicity;
  \item[June 2] Expression of concern from the Lancet for the article about \ac{hcq}, Pr. Raoult launch the hashtag \texttt{\#lancetGate};
  \item[June 5] Retraction of the Lancet article;
  \item[June 24] Pr. Raoult speaks in front of the French Parliament (commission on the management of COVID-19).
\end{description}
In the second period of the debate (after June 2020) the engagement on \ac{hcq} debates is almost monotonically decreasing for the whole period, with an exception on September 2020 influenced by the Gilead affair in which Pr. Raoult shows the conflicts of interest of some member of the Scientific Council with Gilead, the industry producing Remdesivir.

\begin{figure}[!htb]
  \begin{center}
    \includegraphics[width=0.7\linewidth]{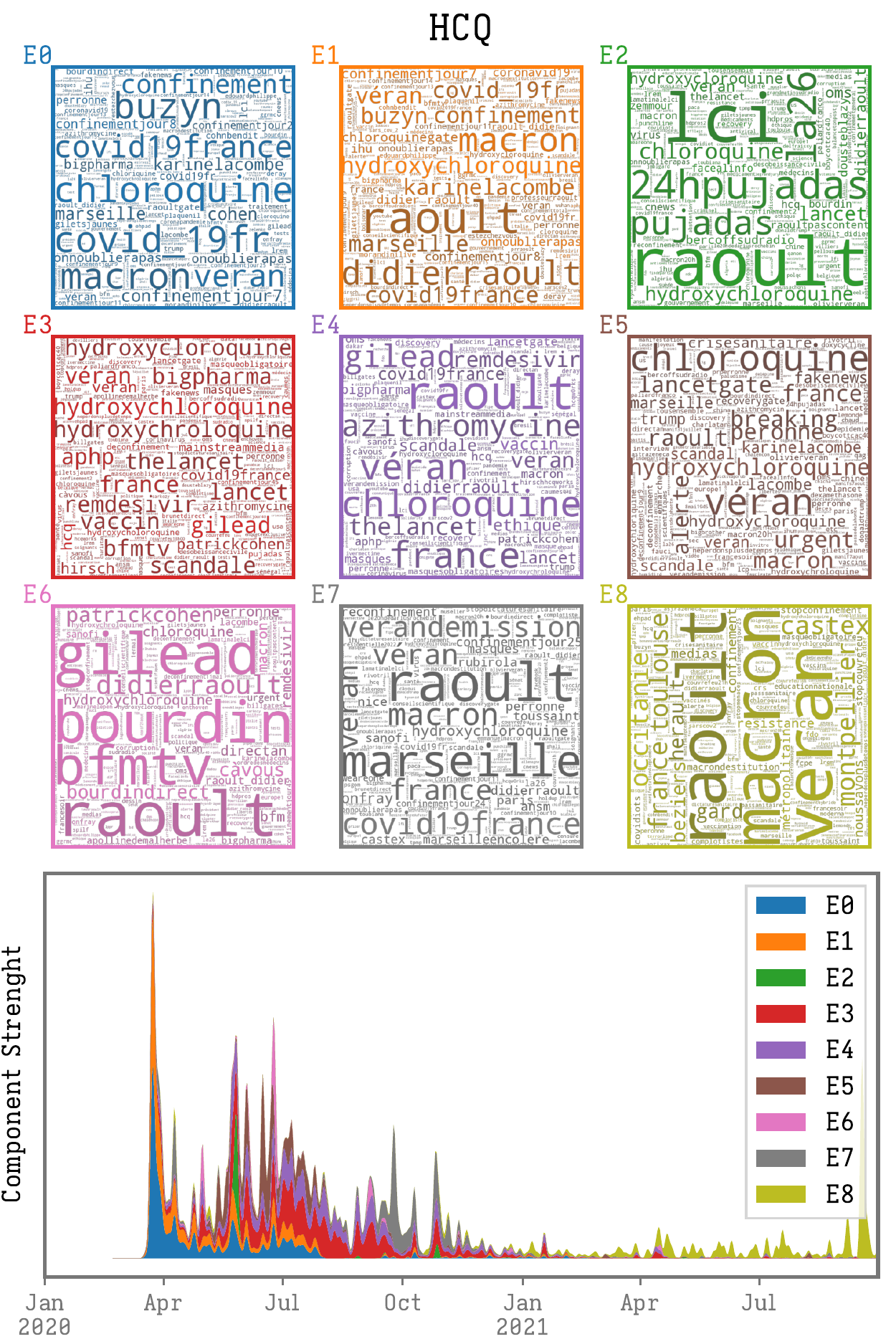}
  \end{center}
  \caption{%
    \csentence{Topics of the online discussion within tweets on \ac{hcq} and related keywords.}
    Word clouds of each component (above).
    Temporal evolution of each component strength (stacked plot below).
  }\label{fig:hcq}
\end{figure}

To better characterize the composition and the timing of the debate, we extracted the subset of tweets focused on \ac{hcq} and related topics and computed the temporal tensor of the hashtags co-occurrence (as described in the methods section).
See Fig.~\ref{fig:hcq}.
This procedure allows to identify the mesoscopic structure of the debate, i.e.\ the different topics debated and the period in which each subject was actively discussed.
The latter can be summarized as follows:
\begin{itemize}
  \item The first set of hashtags, \texttt{E0}, represents the institutional debate on \ac{hcq} which is active from the presentation of the \ac{ijaa} paper until the suspension of the clinical use of \ac{hcq} for COVID-19 treatment;
\item The topic \texttt{E1} is also connected to the initial debate on \ac{hcq} but with a focus on the figure of Pr. Raoult;
\item \texttt{E2} has two clear temporal spikes and is connected to the appearances on television of Pr. Raoult;
\item \texttt{E3}, \texttt{E4} and \texttt{E5} are temporally concentrated between May and September and represent the political debate on the stop to the use of \ac{hcq} for COVID-19 treatments (the institutional debate above all in \texttt{E3}), the lancet gate (\texttt{E5}) and the beginning of the discussion on ``remdesivir'' as the alternative to \ac{hcq} (\texttt{E4});
\item \texttt{E6} concentrates in September has its focus on the Gilead conflicts of interest with the Scientific Council;
\item \texttt{E7} focus on the political figure of Pr. Raoult in Marseille;
\item \texttt{E8} is the late debate in relation with the political system, also showing a peak during the summer, related to the introduction of the sanitary pass.
\end{itemize}

\subsection*{The geography of the \ac{hcq} debate}
In the following we will analyze how the discussion on \ac{hcq} entered the more general Twitter debate on COVID-19 in different French-speaking countries.

A country analysis of the user base of this database reports a heterogeneous distribution of the activity, with most involved users associated to France ($\sim$ 254k) while other countries counting from 1k to 40k users (see Fig.~\ref{fig:stats}, bottom).

The intra- and inter-country flow of information is captured by the hypergraph framework and the random-walk model for the dynamical spreading of information~\cite{faccin2022directedhypergraphs, faccin2022vaccinecritics}.
With this framework we can capture the probability of a single piece of information to be trapped into the original country as in Eq.~\ref{eq:trapping}.

\begin{figure}[!htb]
  \begin{center}
    \includegraphics[width=\linewidth]{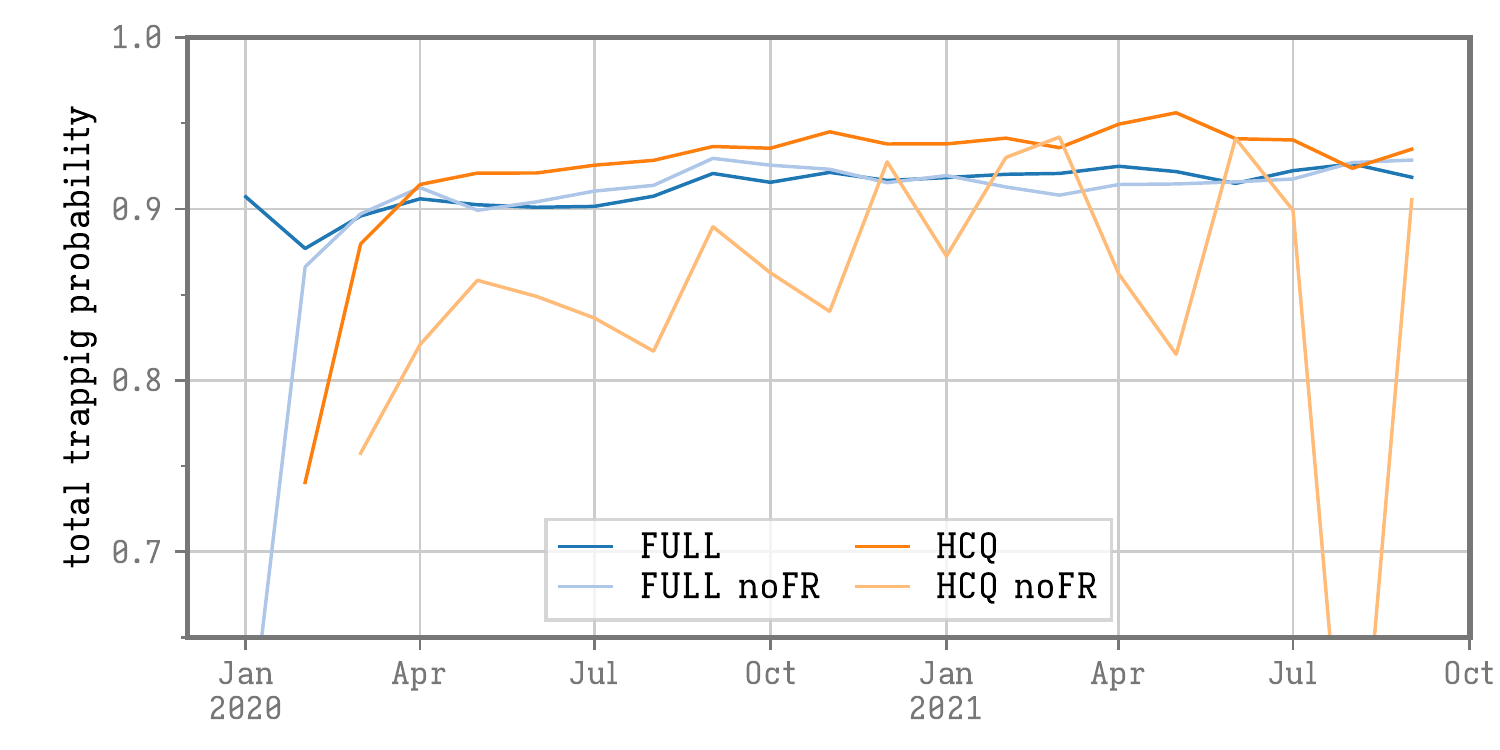}
  \end{center}
  \caption{%
    \csentence{Inward information flow.}
    Amount of information trapped within the country borders on the global hypergraph projected to countries (dark blue) and removing French users (light blue).
    In both cases the probability of being trapped in a country reaches 90\%.
    The same analysis is performed on the subset of tweets on \ac{hcq} (dark and light orange).
    In this case when ignoring the contribution of France the trapping probability drops to 80\% in the first period.
  }\label{fig:inner_flow}
\end{figure}

In Fig.~\ref{fig:inner_flow} we compare the whole debate on COVID-19 and \ac{hcq} along the whole period, to the discussion on \ac{hcq}.
The total trapping probability $p^\circlearrowleft = \sum_C p_C^\circlearrowleft$ form Eq.~\ref{eq:trapping}, is high in both cases, reaching 90\% over the whole time window.
Since users from France account for 52\% of all geotagged users, we removed their tweets and retweets to have a better overview of the international discussion.
Fig.~\ref{fig:inner_flow} reports, with dimmed colors, the total trapping probability for the full dataset and for \ac{hcq} subset, when ignoring France.
In this case, the discussion on \ac{hcq} is slightly less trapped inside the country borders, lowering to around 80--85\% the total trapping probability, in particular in the first period.
This reveals a slightly more international dimension of the discussion on \ac{hcq} on other countries, that is hidden by the huge volume of tweets coming specifically from France (dark orange).

\begin{figure}[!htb]
  \begin{center}
    \includegraphics[width=\linewidth]{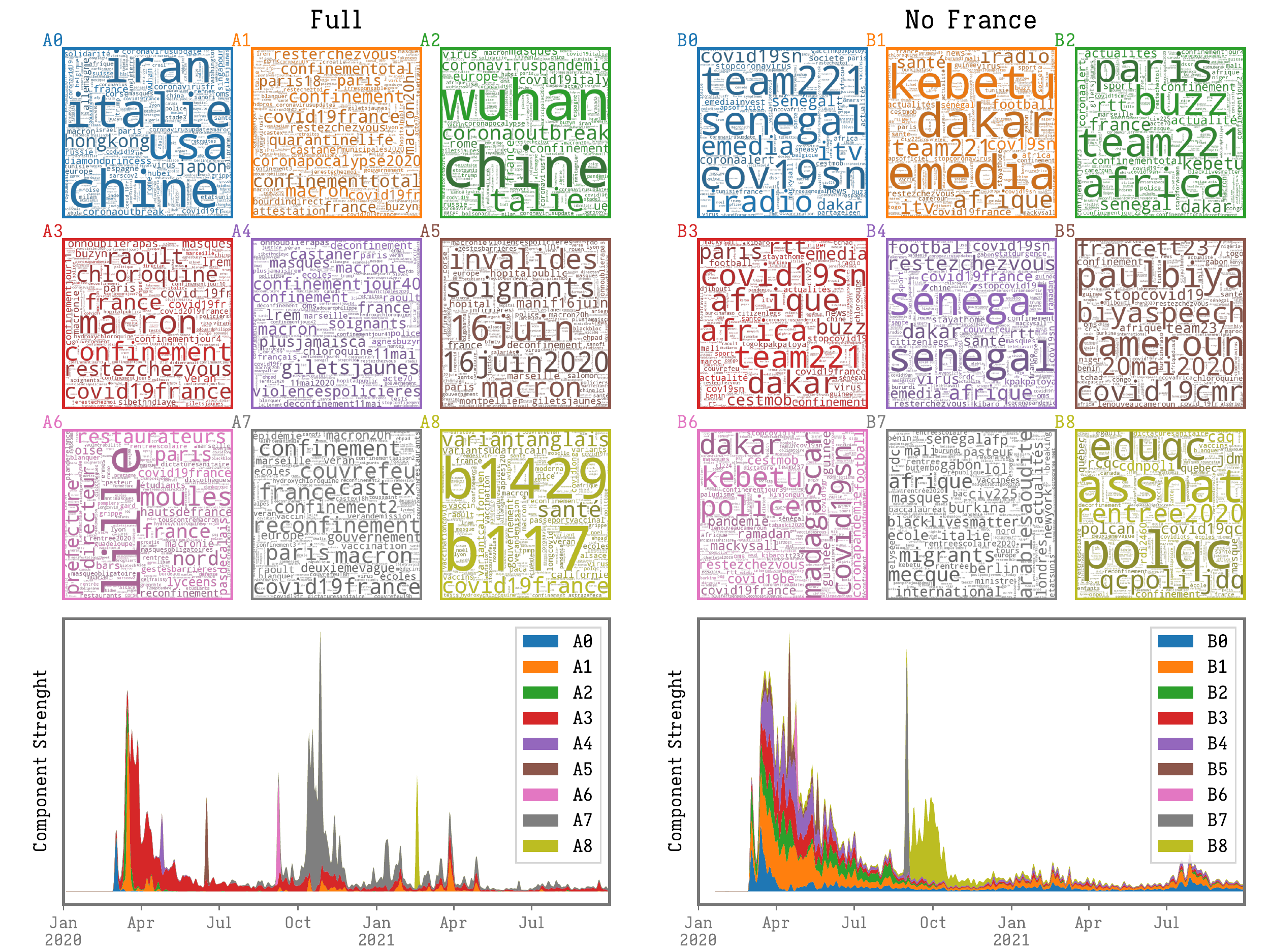}
  \end{center}
  \caption{%
    \csentence{Comparison of topics extracted by \ac{nntf} with (left) and without (right) France.}
    Word clouds of hashtags per component (above, hashtag saturation change is for improved readability).
    The only component that includes \ac{hcq} related hashtags is \texttt{A3}, a long-lasting component in the discussion including French users.
    When considering all countries but France (right), no component has an important contribution to \ac{hcq} topics.
    On the stack plot of component strength (below), some components are temporally limited (linked to events) while others are visible in most of the period of interest (long-lasting topics).
  }\label{fig:nntf_temporal}
\end{figure}

The \ac{nntf} let us extract the components or topics of the temporal hashtag co-occurrence tensor $T$ and their temporal activity.
Fig.~\ref{fig:nntf_temporal} (top) displays the word clouds of each of the 9 components extracted from the tensor, when considering both COVID-19 and \ac{hcq} debates.
The lower part of the figure represents the relative strength of each component.
Note that the most active topic is \texttt{A3} and this is the only one that contains keywords related to \ac{hcq}.
While this topic is active for the whole period, the other topics are limited in time and are mostly related to one-time events (e.g.\ topics \texttt{A0}, \texttt{A1} and \texttt{A2} are related to the onset of the pandemics, \texttt{A8} to the spreading of new virus variants and the other to particular French events).

\begin{figure}[!htb]
  \begin{center}
    \includegraphics[width=\linewidth]{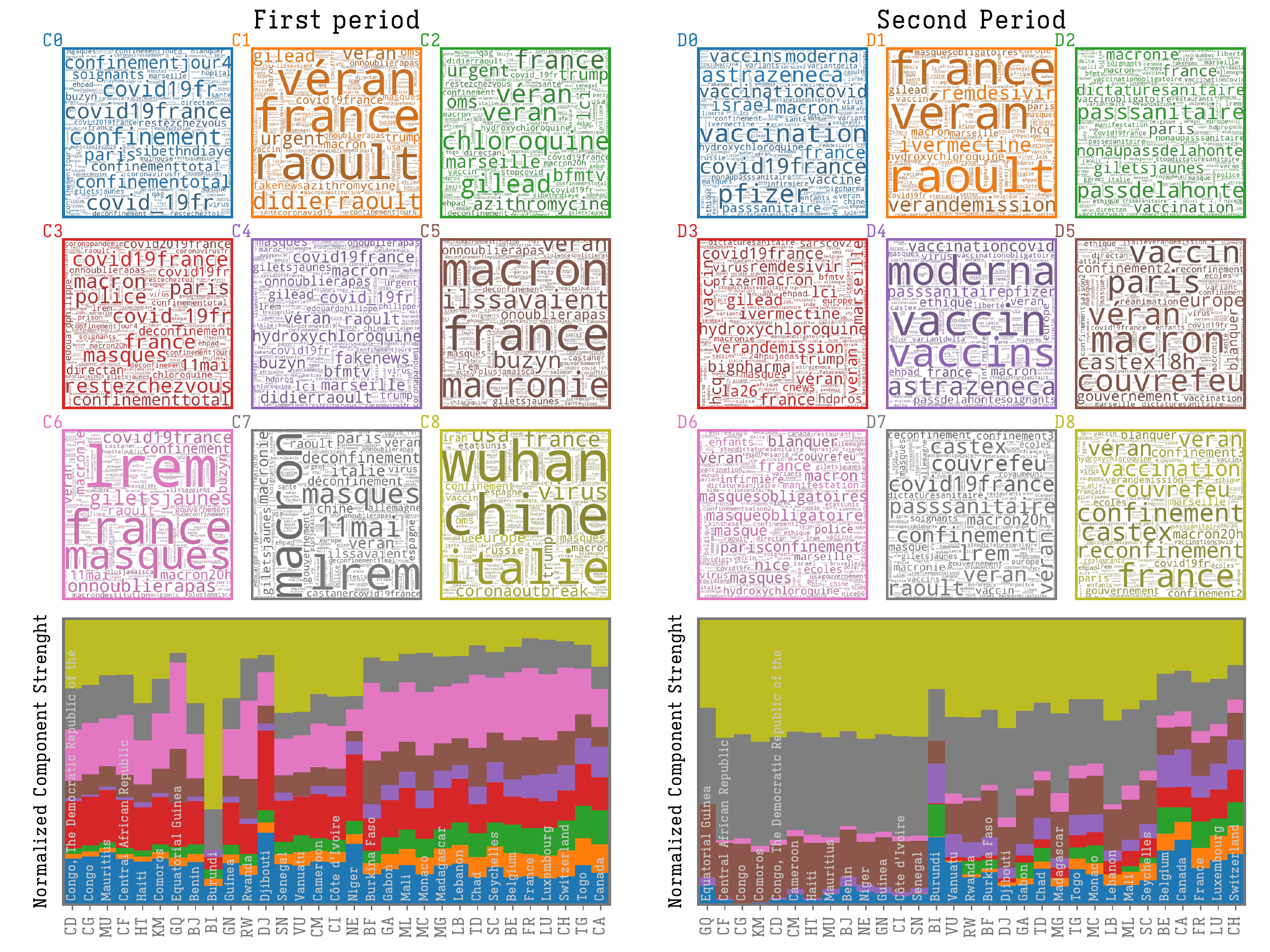}
  \end{center}
  \caption{%
    \csentence{Comparison of topics on the first (left) and second (right) period, per country.}
    Word clouds representing each component (upper plots) show the different distribution of topics in the two periods.
    On the lower plots, a representation of the relative local strength of each component per country is depicted (countries are sorted depending on the inner relative strength of components related to \ac{hcq}: \texttt{C2}, \texttt{C4}, \texttt{D1} and \texttt{D3}).
  }\label{fig:nntf_countries}
\end{figure}

Also in this case, to have a better international view of the topics, we repeated the analysis removing tweets from French users, see the right part of Fig~\ref{fig:nntf_temporal}.
In this framework, while some topics are limited in time as for the full dataset (see \texttt{B6}, \texttt{B7} and \texttt{B8}), others have a persistent activity in time and share a comparable volume of strength.
The latter components are related to country level discussions and in none of the cases there is a reference to a \ac{hcq} keyword.

Similarly, we extract geographic topics by factorizing the country hashtag co-occurrence tensor $K$.
We divide this analysis based on the natural temporal separation mentioned earlier in this section, and the fact that the strength of the \ac{hcq} related components (\texttt{A3} of Fig.~\ref{fig:nntf_temporal} (left)) has higher expression in the first period (up to July 2020).

In the first period (up to July 2020) all components are widely spread among countries, although those related to \ac{hcq} are stronger in European countries (France (FR), Switzerland (CH), Belgium (BE) and Luxembourg (LU)) and Canada, see Fig.~\ref{fig:nntf_countries} (left).
On the bottom of Fig.~\ref{fig:nntf_countries}, the countries are sorted depending on the relative strength of \ac{hcq} components (\texttt{C2} and \texttt{C4}) and European countries and Canada align to the right of the plot.
This effect is even more extreme during the second period (Fig.~\ref{fig:nntf_countries}, right) where the strength of components \texttt{D1} and \texttt{D3} in countries other than Europe and Canada is negligible.
Similarly, we can see the raise of components that attract less attention from the African countries and get localized in the Northern Hemisphere: health pass (\texttt{D2}) and vaccines (\texttt{D4}).

\begin{figure}[!htb]
  \begin{center}
    \includegraphics[width=\linewidth]{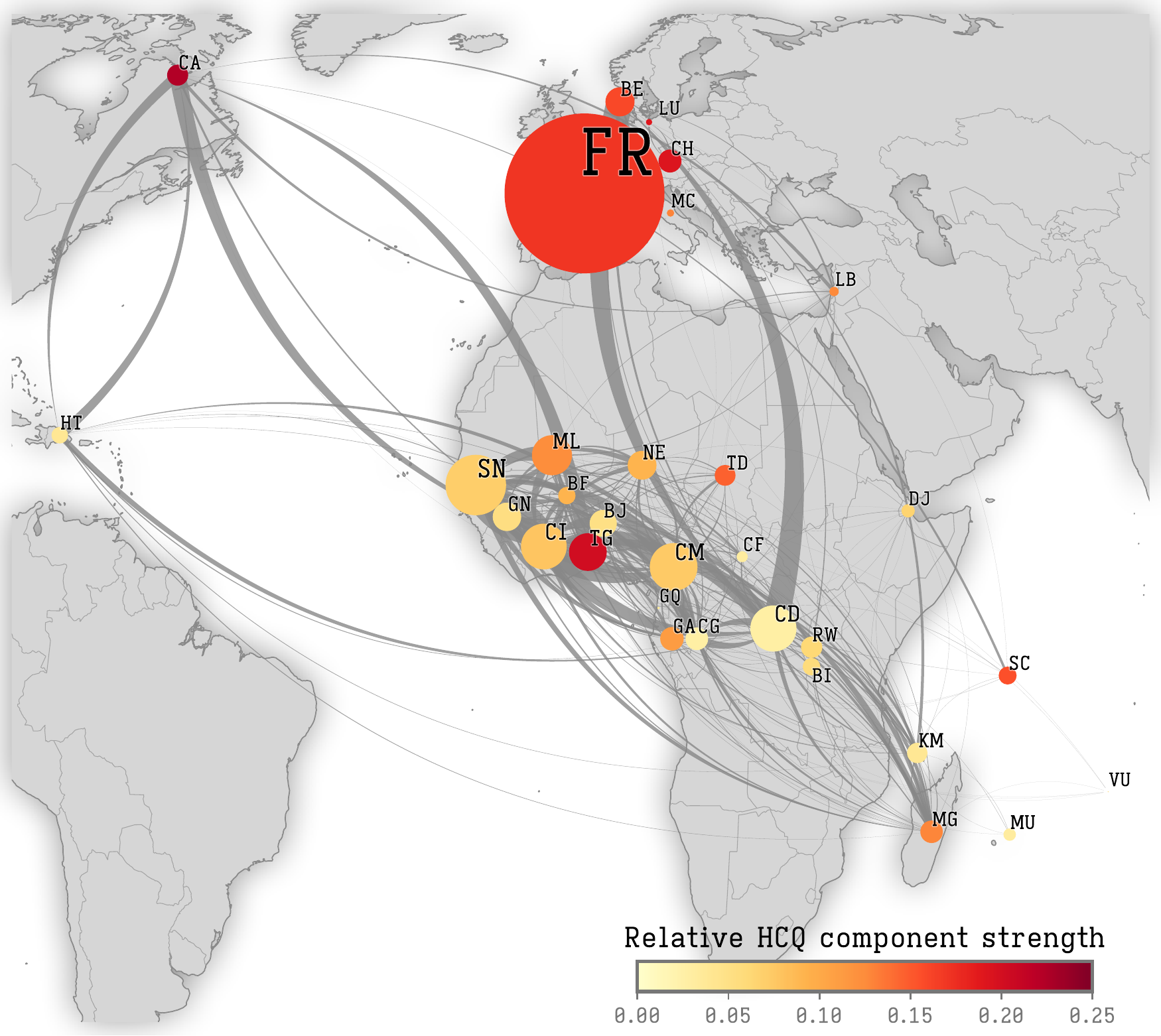}
  \end{center}
  \caption{%
    \csentence{Map of countries connected throughout the online discussion.}
    Link weights correspond to the probability of reaching the neighboring country (links below the average country outgoing probability are hidden for readability).
    Country size is proportional to the square root of the number of users.
    Color codes represent the relative component strength of \texttt{C2} and \texttt{C4} (the components on topics related to \ac{hcq}).
    Note that the components \texttt{C2} and \texttt{C4} localize on the European countries and Canada.
  }\label{fig:nntf_map}
\end{figure}

On Fig.~\ref{fig:nntf_map},  the countries' \ac{hcq} component strength is plotted on a map.
Country colors represent the cumulative strength of component \texttt{C2} and \texttt{C4} in the decomposition of the first period (as in Fig.~\ref{fig:nntf_countries}, left).
This figure highlights how most of the countries where the \ac{hcq} components have higher strength, localize in Europe or North America (expect Togo (TG) which is tightly connected to Canada (CA)).
In this plot link weights are proportional to the probability of the random walker to reach other countries, as in Eq.~\ref{eq:country_prob}.
Links are hidden if their strength is lower than the expected given the origin node strength.

\section*{Discussion}
Debates on \ac{hcq} promptly took an international dimension, fueled by the position taken by prominent figures such as the President of the United States, but also thanks to \acp{ngo} such as the \ac{who}.
However, the management of the epidemic has demonstrated that public health issues are largely country's affaires~\cite{Haldane2021}, and as such the debates that occupy citizens are often on a local or national scale.
National public health decisions will influence and guide the local debate on related topics.
Beyond national borders, countries linked by a common history and areas of similar language use the latter to create a flow of information.

Based on the general dynamics of exchanges around \ac{hcq} on Twitter, we showed the existence of two principal periods: one of very strong engagement between March and July 2020, and then low-level exchanges.
This largely corresponds to the internationalised \enquote{scientific} discussion that has taken place around the publication of results in the main journals and in particular the preliminary results of national trials (DISCOVERY in France, SOLIDARITY by \ac{who} internationally), which are gradually helping to stabilise knowledge on the ineffectiveness of the treatment~\cite{Axfors2021, Gould2021}.
Although debates on \ac{hcq} diluted during the second period as a central theme, they were still a subject of exchanges on Twitter.
In France, this can be explained by a set of public events related to the \ac{hcq}: the vaccination challenge, communications from \ac{ihu} and Pr. Didier Raoult's, notably on Youtube, the legal action against the institute.
In doing so, we draw up a first long chronology of the debates concerning \ac{hcq} and more generally COVID-19 on Twitter, by identifying the main associated themes.
This confirms the pre-eminence of discussions on the political management of the crisis.

Our article also shows the importance of national debates and controversies, beyond the volume of exchange in each country.
Contrary to the impression that discussions on \ac{hcq} have been internationalised around a schedule determined by scientific events, exchanges on Twitter are largely defined by national circulation.
The high degree of recapture of tweets on a national scale indicates a strong nationalisation of the debates.
The slightly lower recapture of \ac{hcq}-specific debates can then be explained not only by the internationalisation of the debates during the first period but also by the importance taken by certain actors in promoting this treatment which has crossed national borders.
However, the results indicate that the very central and media-heavy polemics on \ac{hcq} in France did not have the same impact in other countries.
Thus, while explicitly anti-vaccine positions may have continued to promote \ac{hcq} as a way to maintain alternative course of action~\cite{Bertin2020}, these debates have had no reason to exist in countries where this treatment has become part of the standard management of the disease~\cite{Desclaux2022}.

These results on the national and international circulation around a situation of uncertainty mixing scientific controversy and political debate make it possible, more generally, to inform a geographic approach to digital polemics by relocating the interactions in their national contexts, making it possible to situate the case studies that are often focused on certain countries.
During a global pandemic involving highly differentiated management of the disease, this makes it possible to show that topics controversial and debated in one country are not necessarily so in another.
This work open the possibility to compare national dynamics of controversies, by completing the exchanges in social media with a thick description of national public debates, especially in the legacy media.

Possible biases of this work include the user geotagging based on special keywords that appears in the user profile.
With this approach we may mis-assign users that, while linked to the country they refer to in their profile, are located abroad.
Nevertheless, strong links with the original country drive the user, and most of the conversation happens within likewise users as Fig~\ref{fig:inner_flow} suggests.

We limited our study to the French Twittosphere, being France one of the focal points of the \ac{hcq} discussion, however the results we pointed out in this analysis suggest the need to deepen this study focusing on its internationalization and to the global temporal dimension at the world scale.


\begin{backmatter}

\section*{Declarations}

\subsection*{Availability of data and materials}
The dataset analyzed during the current study (namely the tweet and retweet IDs) are available in the DataCovHCQ repository: \url{https://github.com/maurofaccin/DataCovHCQ}~\cite{DataCovHCQ}.
The same repository contains the Python scripts used for the dataset analysis.

\subsection*{Competing interests}
The authors declare that they have no competing interests.

\subsection*{Funding}
Part of M.F. contribution was supported by Agence Nationale de la Recherche (ANR) through project TRACTRUST, (ANR-20-COVI-0102) and Agence nationale de recherches sur le sida-Maladies infectieuses émergentes (ANRS-MIE), project MEDIACAM-ANRSCOV24.

\subsection*{Authors' contributions}
All authors partecipated to the design of the work.

\subsection*{Acknowledgements}
All authors thank Jeremy K. Ward for fruitful discussion.

\subsection*{Authors' information}
Not applicable

\section*{Abbreviations}
\printacronyms%


\bibliographystyle{bmc-mathphys} 
\bibliography{biblio}      


\begin{thebibliography}{39}
\ifx \bisbn   \undefined \def \bisbn  #1{ISBN #1}\fi
\ifx \binits  \undefined \def \binits#1{#1}\fi
\ifx \bauthor  \undefined \def \bauthor#1{#1}\fi
\ifx \batitle  \undefined \def \batitle#1{#1}\fi
\ifx \bjtitle  \undefined \def \bjtitle#1{#1}\fi
\ifx \bvolume  \undefined \def \bvolume#1{\textbf{#1}}\fi
\ifx \byear  \undefined \def \byear#1{#1}\fi
\ifx \bissue  \undefined \def \bissue#1{#1}\fi
\ifx \bfpage  \undefined \def \bfpage#1{#1}\fi
\ifx \blpage  \undefined \def \blpage #1{#1}\fi
\ifx \burl  \undefined \def \burl#1{\textsf{#1}}\fi
\ifx \doiurl  \undefined \def \doiurl#1{\textsf{#1}}\fi
\ifx \betal  \undefined \def \betal{\textit{et al.}}\fi
\ifx \binstitute  \undefined \def \binstitute#1{#1}\fi
\ifx \binstitutionaled  \undefined \def \binstitutionaled#1{#1}\fi
\ifx \bctitle  \undefined \def \bctitle#1{#1}\fi
\ifx \beditor  \undefined \def \beditor#1{#1}\fi
\ifx \bpublisher  \undefined \def \bpublisher#1{#1}\fi
\ifx \bbtitle  \undefined \def \bbtitle#1{#1}\fi
\ifx \bedition  \undefined \def \bedition#1{#1}\fi
\ifx \bseriesno  \undefined \def \bseriesno#1{#1}\fi
\ifx \blocation  \undefined \def \blocation#1{#1}\fi
\ifx \bsertitle  \undefined \def \bsertitle#1{#1}\fi
\ifx \bsnm \undefined \def \bsnm#1{#1}\fi
\ifx \bsuffix \undefined \def \bsuffix#1{#1}\fi
\ifx \bparticle \undefined \def \bparticle#1{#1}\fi
\ifx \barticle \undefined \def \barticle#1{#1}\fi
\ifx \bconfdate \undefined \def \bconfdate #1{#1}\fi
\ifx \botherref \undefined \def \botherref #1{#1}\fi
\ifx \url \undefined \def \url#1{\textsf{#1}}\fi
\ifx \bchapter \undefined \def \bchapter#1{#1}\fi
\ifx \bbook \undefined \def \bbook#1{#1}\fi
\ifx \bcomment \undefined \def \bcomment#1{#1}\fi
\ifx \oauthor \undefined \def \oauthor#1{#1}\fi
\ifx \citeauthoryear \undefined \def \citeauthoryear#1{#1}\fi
\ifx \endbibitem  \undefined \def \endbibitem {}\fi
\ifx \bconflocation  \undefined \def \bconflocation#1{#1}\fi
\ifx \arxivurl  \undefined \def \arxivurl#1{\textsf{#1}}\fi
\csname PreBibitemsHook\endcsname

\bibitem{Gould2021}
\begin{botherref}
\oauthor{\bsnm{Gould}, \binits{S.}},
\oauthor{\bsnm{Norris}, \binits{S.L.}}:
{Contested effects and chaotic policies: the 2020 story of (hydroxy) chloroquine for treating COVID-19}
(2021).
doi:\doiurl{10.1002/14651858.ED000151}
\end{botherref}
\endbibitem

\bibitem{Schwartz2022}
\begin{barticle}
\bauthor{\bsnm{Schwartz}, \binits{I.S.}},
\bauthor{\bsnm{Boulware}, \binits{D.R.}},
\bauthor{\bsnm{Lee}, \binits{T.C.}}:
\batitle{{Hydroxychloroquine for COVID19: The curtains close on a comedy of errors}}.
\bjtitle{The Lancet Regional Health - Americas}
\bvolume{11},
\bfpage{100268}
(\byear{2022}).
doi:\doiurl{10.1016/j.lana.2022.100268}
\end{barticle}
\endbibitem

\bibitem{Hallin2013}
\begin{barticle}
\bauthor{\bsnm{Hallin}, \binits{D.C.}},
\bauthor{\bsnm{Brandt}, \binits{M.}},
\bauthor{\bsnm{Briggs}, \binits{C.L.}}:
\batitle{{Biomedicalization and the public sphere: Newspaper coverage ofhealth and medicine, 1960s-2000s}}.
\bjtitle{Social Science and Medicine}
\bvolume{96},
\bfpage{121}--\blpage{128}
(\byear{2013}).
doi:\doiurl{10.1016/j.socscimed.2013.07.030}
\end{barticle}
\endbibitem

\bibitem{Gargiulo2020}
\begin{barticle}
\bauthor{\bsnm{Gargiulo}, \binits{F.}},
\bauthor{\bsnm{Cafiero}, \binits{F.}},
\bauthor{\bsnm{Guille-Escuret}, \binits{P.}},
\bauthor{\bsnm{Seror}, \binits{V.}},
\bauthor{\bsnm{Ward}, \binits{J.K.}}:
\batitle{{Asymmetric participation of defenders and critics of vaccines to debates on French-speaking Twitter}}.
\bjtitle{Scientific Reports}
\bvolume{10}(\bissue{1}),
\bfpage{6599}
(\byear{2020}).
doi:\doiurl{10.1038/s41598-020-62880-5}.
\arxivurl{1909.08311}
\end{barticle}
\endbibitem

\bibitem{Tuccori2020}
\begin{barticle}
\bauthor{\bsnm{Tuccori}, \binits{M.}},
\bauthor{\bsnm{Convertino}, \binits{I.}},
\bauthor{\bsnm{Ferraro}, \binits{S.}},
\bauthor{\bsnm{Cappello}, \binits{E.}},
\bauthor{\bsnm{Valdiserra}, \binits{G.}},
\bauthor{\bsnm{Focosi}, \binits{D.}},
\bauthor{\bsnm{Blandizzi}, \binits{C.}}:
\batitle{{The Impact of the COVID-19 ``Infodemic'' on Drug-Utilization Behaviors: Implications for Pharmacovigilance}}.
\bjtitle{Drug Safety}
\bvolume{43}(\bissue{8}),
\bfpage{699}--\blpage{709}
(\byear{2020}).
doi:\doiurl{10.1007/s40264-020-00965-w}
\end{barticle}
\endbibitem

\bibitem{Rughinis2020}
\begin{barticle}
\bauthor{\bsnm{Rughiniş}, \binits{C.}},
\bauthor{\bsnm{Dima}, \binits{L.}},
\bauthor{\bsnm{Vasile}, \binits{S.}}:
\batitle{{Hydroxychloroquine and COVID-19: Lack of Efficacy and the Social Construction of Plausibility}}.
\bjtitle{American Journal of Therapeutics}
\bvolume{27}(\bissue{6}),
\bfpage{573}--\blpage{583}
(\byear{2020}).
doi:\doiurl{10.1097/MJT.0000000000001294}
\end{barticle}
\endbibitem

\bibitem{Mutlu2020}
\begin{barticle}
\bauthor{\bsnm{Mutlu}, \binits{E.C.}},
\bauthor{\bsnm{Oghaz}, \binits{T.}},
\bauthor{\bsnm{Jasser}, \binits{J.}},
\bauthor{\bsnm{Tutunculer}, \binits{E.}},
\bauthor{\bsnm{Rajabi}, \binits{A.}},
\bauthor{\bsnm{Tayebi}, \binits{A.}},
\bauthor{\bsnm{Ozmen}, \binits{O.}},
\bauthor{\bsnm{Garibay}, \binits{I.}}:
\batitle{{A stance data set on polarized conversations on Twitter about the efficacy of hydroxychloroquine as a treatment for COVID-19}}.
\bjtitle{Data in Brief}
\bvolume{33},
\bfpage{106401}
(\byear{2020}).
doi:\doiurl{10.1016/j.dib.2020.106401}.
\arxivurl{2009.01188}
\end{barticle}
\endbibitem

\bibitem{Pearson2021}
\begin{barticle}
\bauthor{\bsnm{Pearson}, \binits{H.}}:
\batitle{{How COVID broke the evidence pipeline}}.
\bjtitle{Nature}
\bvolume{593}(\bissue{7858}),
\bfpage{182}--\blpage{185}
(\byear{2021}).
doi:\doiurl{10.1038/d41586-021-01246-x}
\end{barticle}
\endbibitem

\bibitem{Schultz2021a}
\begin{barticle}
\bauthor{\bsnm{Schultz}, \binits{{\'{E}}.}},
\bauthor{\bsnm{Ward}, \binits{J.K.}}:
\batitle{{Science under Covid-19's magnifying glass: Lessons from the first months of the chloroquine debate in the French press}}.
\bjtitle{Journal of Sociology}
\bvolume{58}(\bissue{1}),
\bfpage{76}--\blpage{94}
(\byear{2022}).
doi:\doiurl{10.1177/1440783321999453}
\end{barticle}
\endbibitem

\bibitem{Marcon2021}
\begin{barticle}
\bauthor{\bsnm{Marcon}, \binits{A.R.}},
\bauthor{\bsnm{Caulfield}, \binits{T.}}:
\batitle{{The Hydroxychloroquine Twitter War: A case study examining polarization in science communication}}.
\bjtitle{First Monday}
(\byear{2021}).
doi:\doiurl{10.5210/fm.v26i10.11707}
\end{barticle}
\endbibitem

\bibitem{Origgi2022}
\begin{botherref}
\oauthor{\bsnm{Origgi}, \binits{G.}},
\oauthor{\bsnm{Branch-smith}, \binits{T.}},
\oauthor{\bsnm{Trust}, \binits{T.M.}}:
{Why Trust Raoult? How Social Indicators Inform the Reputations of Experts}.
Social Epistemology
(2022)
\end{botherref}
\endbibitem

\bibitem{Krause2022}
\begin{barticle}
\bauthor{\bsnm{Krause}, \binits{N.M.}},
\bauthor{\bsnm{Freiling}, \binits{I.}},
\bauthor{\bsnm{Scheufele}, \binits{D.A.}}:
\batitle{{The ``Infodemic'' Infodemic: Toward a More Nuanced Understanding of Truth-Claims and the Need for (Not) Combatting Misinformation}}.
\bjtitle{Annals of the American Academy of Political and Social Science}
\bvolume{700}(\bissue{1}),
\bfpage{112}--\blpage{123}
(\byear{2022}).
doi:\doiurl{10.1177/00027162221086263}
\end{barticle}
\endbibitem

\bibitem{Ektorp2020}
\begin{barticle}
\bauthor{\bsnm{Ektorp}, \binits{E.}}:
\batitle{{Death threats after a trial on chloroquine for COVID-19}}.
\bjtitle{The Lancet. Infectious diseases}
\bvolume{20}(\bissue{6}),
\bfpage{661}
(\byear{2020}).
doi:\doiurl{10.1016/S1473-3099(20)30383-2}
\end{barticle}
\endbibitem

\bibitem{Berlivet2020}
\begin{barticle}
\bauthor{\bsnm{Berlivet}, \binits{L.}},
\bauthor{\bsnm{L{\"{o}}wy}, \binits{I.}}:
\batitle{{Hydroxychloroquine Controversies: Clinical Trials, Epistemology, and the Democratization of Science}}.
\bjtitle{Medical Anthropology Quarterly}
\bvolume{34}(\bissue{4}),
\bfpage{525}--\blpage{541}
(\byear{2020}).
doi:\doiurl{10.1111/maq.12622}
\end{barticle}
\endbibitem

\bibitem{Schultz2022a}
\begin{barticle}
\bauthor{\bsnm{Schultz}, \binits{{\'{E}}.}},
\bauthor{\bsnm{Atlani-Duault}, \binits{L.}},
\bauthor{\bsnm{Peretti-Watel}, \binits{P.}},
\bauthor{\bsnm{Ward}, \binits{J.K.}}:
\batitle{{Does the public know when a scientific controversy is over? Public perceptions of hydroxychloroquine in France between April 2020 and June 2021}}.
\bjtitle{Therapies}
(\bissue{January})
(\byear{2022}).
doi:\doiurl{10.1016/j.therap.2022.01.008}
\end{barticle}
\endbibitem

\bibitem{Smyrnaios2021}
\begin{barticle}
\bauthor{\bsnm{Smyrnaios}, \binits{N.}},
\bauthor{\bsnm{Tsimboukis}, \binits{P.}},
\bauthor{\bsnm{Loub{\`{e}}re}, \binits{L.}}:
\batitle{{La controverse autour de Didier Raoult et de sa proposition th{\'{e}}rapeutique contre le Covid-19 sur Twitter : analyse de r{\'{e}}seaux et de discours}}.
\bjtitle{Communiquer. Revue de communication sociale et publique}
\bvolume{31},
\bfpage{1}--\blpage{20}
(\byear{2021})
\end{barticle}
\endbibitem

\bibitem{Dubois2021}
\begin{bchapter}
\bauthor{\bsnm{Dubois}, \binits{M.}},
\bauthor{\bsnm{Frenod-Dunaud}, \binits{A.}},
\bauthor{\bsnm{Gargiulo}, \binits{F.}},
\bauthor{\bsnm{Guaspare-Carton}, \binits{C.}}:
\bctitle{Le contrôle par les pairs au temps du coronavirus}.
In: \beditor{\bsnm{Chauvin}, \binits{P.M.}},
\beditor{\bsnm{Clement}, \binits{A.}} (eds.)
\bbtitle{Sorbonnavirus, Regards sur la Crise du Coronavirus}.
\bpublisher{Sorbonne Université Presses}, \blocation{???}
(\byear{2021})
\end{bchapter}
\endbibitem

\bibitem{Bharti2022}
\begin{botherref}
\oauthor{\bsnm{Bharti}, \binits{N.}},
\oauthor{\bsnm{Sismondo}, \binits{S.}}:
{Political Prescriptions: Three Pandemic Stories}.
Science, Technology \& Human Values,
016224392211238
(2022).
doi:\doiurl{10.1177/01622439221123831}
\end{botherref}
\endbibitem

\bibitem{Desclaux2022}
\begin{botherref}
\oauthor{\bsnm{Desclaux}, \binits{A.}}:
{Covid-19: En Afrique de l'Ouest, le vaccin n'est pas le nouveau ``magic bullet''}
(2022)
\end{botherref}
\endbibitem

\bibitem{Axfors2021}
\begin{barticle}
\bauthor{\bsnm{Axfors}, \binits{C.}},
\bauthor{\bsnm{Schmitt}, \binits{A.M.}},
\bauthor{\bsnm{Janiaud}, \binits{P.}},
\bauthor{\bparticle{van't} \bsnm{Hooft}, \binits{J.}},
\bauthor{\bsnm{Abd-Elsalam}, \binits{S.}},
\bauthor{\bsnm{Abdo}, \binits{E.F.}},
\bauthor{\bsnm{Abella}, \binits{B.S.}},
\bauthor{\bsnm{Akram}, \binits{J.}},
\bauthor{\bsnm{Amaravadi}, \binits{R.K.}},
\bauthor{\bsnm{Angus}, \binits{D.C.}},
\bauthor{\bsnm{Arabi}, \binits{Y.M.}},
\bauthor{\bsnm{Azhar}, \binits{S.}},
\bauthor{\bsnm{Baden}, \binits{L.R.}},
\bauthor{\bsnm{Baker}, \binits{A.W.}},
\bauthor{\bsnm{Belkhir}, \binits{L.}},
\bauthor{\bsnm{Benfield}, \binits{T.}},
\bauthor{\bsnm{Berrevoets}, \binits{M.A.H.}},
\bauthor{\bsnm{Chen}, \binits{C.P.}},
\bauthor{\bsnm{Chen}, \binits{T.C.}},
\bauthor{\bsnm{Cheng}, \binits{S.H.}},
\bauthor{\bsnm{Cheng}, \binits{C.Y.}},
\bauthor{\bsnm{Chung}, \binits{W.S.}},
\bauthor{\bsnm{Cohen}, \binits{Y.Z.}},
\bauthor{\bsnm{Cowan}, \binits{L.N.}},
\bauthor{\bsnm{Dalgard}, \binits{O.}},
\bauthor{\bsnm{{de Almeida e Val}}, \binits{F.F.}},
\bauthor{\bparticle{de} \bsnm{Lacerda}, \binits{M.V.G.}},
\bauthor{\bparticle{de} \bsnm{Melo}, \binits{G.C.}},
\bauthor{\bsnm{Derde}, \binits{L.}},
\bauthor{\bsnm{Dubee}, \binits{V.}},
\bauthor{\bsnm{Elfakir}, \binits{A.}},
\bauthor{\bsnm{Gordon}, \binits{A.C.}},
\bauthor{\bsnm{Hernandez-Cardenas}, \binits{C.M.}},
\bauthor{\bsnm{Hills}, \binits{T.}},
\bauthor{\bsnm{Hoepelman}, \binits{A.I.M.}},
\bauthor{\bsnm{Huang}, \binits{Y.W.}},
\bauthor{\bsnm{Igau}, \binits{B.}},
\bauthor{\bsnm{Jin}, \binits{R.}},
\bauthor{\bsnm{Jurado-Camacho}, \binits{F.}},
\bauthor{\bsnm{Khan}, \binits{K.S.}},
\bauthor{\bsnm{Kremsner}, \binits{P.G.}},
\bauthor{\bsnm{Kreuels}, \binits{B.}},
\bauthor{\bsnm{Kuo}, \binits{C.Y.}},
\bauthor{\bsnm{Le}, \binits{T.}},
\bauthor{\bsnm{Lin}, \binits{Y.C.}},
\bauthor{\bsnm{Lin}, \binits{W.P.}},
\bauthor{\bsnm{Lin}, \binits{T.H.}},
\bauthor{\bsnm{Lyngbakken}, \binits{M.N.}},
\bauthor{\bsnm{McArthur}, \binits{C.}},
\bauthor{\bsnm{McVerry}, \binits{B.J.}},
\bauthor{\bsnm{Meza-Meneses}, \binits{P.}},
\bauthor{\bsnm{Monteiro}, \binits{W.M.}},
\bauthor{\bsnm{Morpeth}, \binits{S.C.}},
\bauthor{\bsnm{Mourad}, \binits{A.}},
\bauthor{\bsnm{Mulligan}, \binits{M.J.}},
\bauthor{\bsnm{Murthy}, \binits{S.}},
\bauthor{\bsnm{Naggie}, \binits{S.}},
\bauthor{\bsnm{Narayanasamy}, \binits{S.}},
\bauthor{\bsnm{Nichol}, \binits{A.}},
\bauthor{\bsnm{Novack}, \binits{L.A.}},
\bauthor{\bsnm{O'Brien}, \binits{S.M.}},
\bauthor{\bsnm{Okeke}, \binits{N.L.}},
\bauthor{\bsnm{Perez}, \binits{L.}},
\bauthor{\bsnm{Perez-Padilla}, \binits{R.}},
\bauthor{\bsnm{Perrin}, \binits{L.}},
\bauthor{\bsnm{Remigio-Luna}, \binits{A.}},
\bauthor{\bsnm{Rivera-Martinez}, \binits{N.E.}},
\bauthor{\bsnm{Rockhold}, \binits{F.W.}},
\bauthor{\bsnm{Rodriguez-Llamazares}, \binits{S.}},
\bauthor{\bsnm{Rolfe}, \binits{R.}},
\bauthor{\bsnm{Rosa}, \binits{R.}},
\bauthor{\bsnm{R{\o}sj{\o}}, \binits{H.}},
\bauthor{\bsnm{Sampaio}, \binits{V.S.}},
\bauthor{\bsnm{Seto}, \binits{T.B.}},
\bauthor{\bsnm{Shehzad}, \binits{M.}},
\bauthor{\bsnm{Soliman}, \binits{S.}},
\bauthor{\bsnm{Stout}, \binits{J.E.}},
\bauthor{\bsnm{Thirion-Romero}, \binits{I.}},
\bauthor{\bsnm{Troxel}, \binits{A.B.}},
\bauthor{\bsnm{Tseng}, \binits{T.Y.}},
\bauthor{\bsnm{Turner}, \binits{N.A.}},
\bauthor{\bsnm{Ulrich}, \binits{R.J.}},
\bauthor{\bsnm{Walsh}, \binits{S.R.}},
\bauthor{\bsnm{Webb}, \binits{S.A.}},
\bauthor{\bsnm{Weehuizen}, \binits{J.M.}},
\bauthor{\bsnm{Velinova}, \binits{M.}},
\bauthor{\bsnm{Wong}, \binits{H.L.}},
\bauthor{\bsnm{Wrenn}, \binits{R.}},
\bauthor{\bsnm{Zampieri}, \binits{F.G.}},
\bauthor{\bsnm{Zhong}, \binits{W.}},
\bauthor{\bsnm{Moher}, \binits{D.}},
\bauthor{\bsnm{Goodman}, \binits{S.N.}},
\bauthor{\bsnm{Ioannidis}, \binits{J.P.A.}},
\bauthor{\bsnm{Hemkens}, \binits{L.G.}}:
\batitle{{Mortality outcomes with hydroxychloroquine and chloroquine in COVID-19 from an international collaborative meta-analysis of randomized trials}}.
\bjtitle{Nature Communications}
\bvolume{12}(\bissue{1}),
\bfpage{1}--\blpage{13}
(\byear{2021}).
doi:\doiurl{10.1038/s41467-021-22446-z}
\end{barticle}
\endbibitem

\bibitem{Taccone2022}
\begin{barticle}
\bauthor{\bsnm{Taccone}, \binits{F.S.}},
\bauthor{\bsnm{Hites}, \binits{M.}},
\bauthor{\bsnm{Dauby}, \binits{N.}}:
\batitle{{From hydroxychloroquine to ivermectin: how unproven ``cures'' can go viral}}.
\bjtitle{Clinical Microbiology and Infection}
\bvolume{28}(\bissue{4}),
\bfpage{472}--\blpage{474}
(\byear{2022}).
doi:\doiurl{10.1016/j.cmi.2022.01.008}
\end{barticle}
\endbibitem

\bibitem{Casaroes2021}
\begin{barticle}
\bauthor{\bsnm{Casar{\~{o}}es}, \binits{G.}},
\bauthor{\bsnm{Magalh{\~{a}}es}, \binits{D.}}:
\batitle{{The hydroxychloroquine alliance: how far-right leaders and alt-science preachers came together to promote a miracle drug}}.
\bjtitle{Revista de Administracao Publica}
\bvolume{55}(\bissue{1}),
\bfpage{197}--\blpage{214}
(\byear{2021}).
doi:\doiurl{10.1590/0034-761220200556}
\end{barticle}
\endbibitem

\bibitem{Osuagwu2021}
\begin{barticle}
\bauthor{\bsnm{Osuagwu}, \binits{U.L.}},
\bauthor{\bsnm{Nwaeze}, \binits{O.}},
\bauthor{\bsnm{Ovenseri-Ogbomo}, \binits{G.}},
\bauthor{\bsnm{Oloruntoba}, \binits{R.}},
\bauthor{\bsnm{Ekpenyong}, \binits{B.}},
\bauthor{\bsnm{Mashige}, \binits{K.P.}},
\bauthor{\bsnm{Timothy}, \binits{C.}},
\bauthor{\bsnm{Ishaya}, \binits{T.}},
\bauthor{\bsnm{Langsi}, \binits{R.}},
\bauthor{\bsnm{Charwe}, \binits{D.}},
\bauthor{\bsnm{Abu}, \binits{E.K.}},
\bauthor{\bsnm{Chundung}, \binits{M.A.}},
\bauthor{\bsnm{Agho}, \binits{K.E.}}:
\batitle{{Opinion and uptake of chloroquine for treatment of COVID-19 during the mandatory lockdown in the sub-Saharan African region}}.
\bjtitle{African Journal of Primary Health Care \& Family Medicine}
\bvolume{13}(\bissue{1}),
\bfpage{1}--\blpage{8}
(\byear{2021}).
doi:\doiurl{10.4102/phcfm.v13i1.2795}
\end{barticle}
\endbibitem

\bibitem{Mukankubito2021}
\begin{botherref}
\oauthor{\bsnm{Mukankubito}, \binits{I.}},
\oauthor{\bsnm{Annan}, \binits{E.A.}},
\oauthor{\bsnm{Sougou}, \binits{A.}},
\oauthor{\bsnm{Taguembou}, \binits{D.}},
\oauthor{\bsnm{Kniazkov}, \binits{S.}},
\oauthor{\bsnm{Loua}, \binits{A.}},
\oauthor{\bsnm{Mankele}, \binits{R.}},
\oauthor{\bsnm{Nikiema}, \binits{J.-B.}},
\oauthor{\bsnm{Bisoborwa}, \binits{G.}},
\oauthor{\bsnm{Julius}, \binits{O.K.M.}}:
{COVID-19 Treatment Protocols in the WHO African Region - Results of a Survey},
1--19
(2021)
\end{botherref}
\endbibitem

\bibitem{Atlani-Duault2016}
\begin{barticle}
\bauthor{\bsnm{Atlani-Duault}, \binits{L.}},
\bauthor{\bsnm{Dozon}, \binits{J.P.}},
\bauthor{\bsnm{Wilson}, \binits{A.}},
\bauthor{\bsnm{Delfraissy}, \binits{J.F.}},
\bauthor{\bsnm{Moatti}, \binits{J.P.}}:
\batitle{{State humanitarian verticalism versus universal health coverage: A century of French international health assistance revisited}}.
\bjtitle{The Lancet}
\bvolume{387}(\bissue{10034}),
\bfpage{2250}--\blpage{2262}
(\byear{2016}).
doi:\doiurl{10.1016/S0140-6736(16)00379-2}
\end{barticle}
\endbibitem

\bibitem{Lapostolle2021}
\begin{barticle}
\bauthor{\bsnm{Lapostolle}, \binits{F.}},
\bauthor{\bsnm{Vianu}, \binits{I.}},
\bauthor{\bsnm{{De Stefano}}, \binits{C.}},
\bauthor{\bsnm{Goix}, \binits{L.}},
\bauthor{\bsnm{Petrovic}, \binits{T.}},
\bauthor{\bsnm{Adnet}, \binits{F.}}:
\batitle{{COVID-19 Epidemic: Chloroquine, a French Obsession?}}
\bjtitle{La Presse M{\'{e}}dicale Open}
\bvolume{2},
\bfpage{100007}
(\byear{2021}).
doi:\doiurl{10.1016/j.lpmope.2021.100007}
\end{barticle}
\endbibitem

\bibitem{Roy2020}
\begin{barticle}
\bauthor{\bsnm{Roy}, \binits{M.}},
\bauthor{\bsnm{Moreau}, \binits{N.}},
\bauthor{\bsnm{Rousseau}, \binits{C.}},
\bauthor{\bsnm{Mercier}, \binits{A.}},
\bauthor{\bsnm{Wilson}, \binits{A.}},
\bauthor{\bsnm{Atlani-Duault}, \binits{L.}}:
\batitle{{Ebola and Localized Blame on Social Media: Analysis of Twitter and Facebook Conversations During the 2014--2015 Ebola Epidemic}}.
\bjtitle{Culture, Medicine and Psychiatry}
\bvolume{44}(\bissue{1}),
\bfpage{56}--\blpage{79}
(\byear{2020}).
doi:\doiurl{10.1007/s11013-019-09635-8}
\end{barticle}
\endbibitem

\bibitem{gazu}
\begin{botherref}
\oauthor{\bsnm{Ooghe-Tabanou}, \binits{B.}},
\oauthor{\bsnm{Farjas}, \binits{J.}},
\oauthor{\bsnm{Mazoyer}, \binits{B.}}:
Gazouilloire, Twitter stream + search API grabber.
\url{https://github.com/medialab/gazouilloire}
\end{botherref}
\endbibitem

\bibitem{DataCovVac}
\begin{botherref}
\oauthor{\bsnm{Faccin}, \binits{M.}}:
maurofaccin/DataCovVac.
doi:\doiurl{10.5281/zenodo.7870249}.
\url{https://doi.org/10.5281/zenodo.7870249}
\end{botherref}
\endbibitem

\bibitem{faccin2022vaccinecritics}
\begin{botherref}
\oauthor{\bsnm{Faccin}, \binits{M.}},
\oauthor{\bsnm{Gargiulo}, \binits{F.}},
\oauthor{\bsnm{Atlani-Duault}, \binits{L.}},
\oauthor{\bsnm{Ward}, \binits{J.K.}}:
Assessing the influence of french vaccine critics during the two first years of the { COVID}-19 pandemic.
arxiv:2202.10952
(2022).
\arxivurl{2202.10952}
\end{botherref}
\endbibitem

\bibitem{DataCovHCQ}
\begin{botherref}
\oauthor{\bsnm{Faccin}, \binits{M.}}:
maurofaccin/DataCovHCQ.
doi:\doiurl{10.5281/zenodo.7870286}.
\url{https://doi.org/10.5281/zenodo.7870286}
\end{botherref}
\endbibitem

\bibitem{Ausiello_2017}
\begin{barticle}
\bauthor{\bsnm{Ausiello}, \binits{G.}},
\bauthor{\bsnm{Laura}, \binits{L.}}:
\batitle{Directed hypergraphs: Introduction and fundamental algorithms{\textemdash}a survey}.
\bjtitle{Theoretical Computer Science}
\bvolume{658},
\bfpage{293}--\blpage{306}
(\byear{2017}).
doi:\doiurl{10.1016/j.tcs.2016.03.016}
\end{barticle}
\endbibitem

\bibitem{faccin2022directedhypergraphs}
\begin{botherref}
\oauthor{\bsnm{Faccin}, \binits{M.}}:
Measuring dynamical systems on directed hypergraphs.
arxiv:2202.12810
(2022).
\arxivurl{2202.12810}
\end{botherref}
\endbibitem

\bibitem{Shashua_2005_nntf}
\begin{bchapter}
\bauthor{\bsnm{Shashua}, \binits{A.}},
\bauthor{\bsnm{Hazan}, \binits{T.}}:
\bctitle{Non-negative tensor factorization with applications to statistics and computer vision}.
In: \bbtitle{Proceedings of the 22nd International Conference on Machine Learning - {ICML} 2005}.
\bpublisher{{ACM} Press}, \blocation{???}
(\byear{2005}).
doi:\doiurl{10.1145/1102351.1102451}.
\burl{https://doi.org/10.1145/1102351.1102451}
\end{bchapter}
\endbibitem

\bibitem{Cichocki_2009_nn_factorization}
\begin{bbook}
\bauthor{\bsnm{Cichocki}, \binits{A.}},
\bauthor{\bsnm{Zdunek}, \binits{R.}},
\bauthor{\bsnm{Phan}, \binits{A.H.}},
\bauthor{\bsnm{Amari}, \binits{S.-I.}}:
\bbtitle{Nonnegative Matrix and Tensor Factorizations}.
\bpublisher{John Wiley {\&} Sons, Ltd}, \blocation{???}
(\byear{2009}).
doi:\doiurl{10.1002/9780470747278}.
\burl{https://doi.org/10.1002/9780470747278}
\end{bbook}
\endbibitem

\bibitem{M_rup_2011_application_factorizations}
\begin{barticle}
\bauthor{\bsnm{M{\o}rup}, \binits{M.}}:
\batitle{Applications of tensor (multiway array) factorizations and decompositions in data mining}.
\bjtitle{WIREs Data Mining and Knowledge Discovery}
\bvolume{1}(\bissue{1}),
\bfpage{24}--\blpage{40}
(\byear{2011}).
doi:\doiurl{10.1002/widm.1}
\end{barticle}
\endbibitem

\bibitem{Gauvin_2014_nntf_communities}
\begin{barticle}
\bauthor{\bsnm{Gauvin}, \binits{L.}},
\bauthor{\bsnm{Panisson}, \binits{A.}},
\bauthor{\bsnm{Cattuto}, \binits{C.}}:
\batitle{Detecting the community structure and activity patterns of temporal networks: A non-negative tensor factorization approach}.
\bjtitle{PLoS {ONE}}
\bvolume{9}(\bissue{1}),
\bfpage{86028}
(\byear{2014}).
doi:\doiurl{10.1371/journal.pone.0086028}
\end{barticle}
\endbibitem

\bibitem{Haldane2021}
\begin{barticle}
\bauthor{\bsnm{Haldane}, \binits{V.}},
\bauthor{\bsnm{{De Foo}}, \binits{C.}},
\bauthor{\bsnm{Abdalla}, \binits{S.M.}},
\bauthor{\bsnm{Jung}, \binits{A.S.}},
\bauthor{\bsnm{Tan}, \binits{M.}},
\bauthor{\bsnm{Wu}, \binits{S.}},
\bauthor{\bsnm{Chua}, \binits{A.}},
\bauthor{\bsnm{Verma}, \binits{M.}},
\bauthor{\bsnm{Shrestha}, \binits{P.}},
\bauthor{\bsnm{Singh}, \binits{S.}},
\bauthor{\bsnm{Perez}, \binits{T.}},
\bauthor{\bsnm{Tan}, \binits{S.M.}},
\bauthor{\bsnm{Bartos}, \binits{M.}},
\bauthor{\bsnm{Mabuchi}, \binits{S.}},
\bauthor{\bsnm{Bonk}, \binits{M.}},
\bauthor{\bsnm{McNab}, \binits{C.}},
\bauthor{\bsnm{Werner}, \binits{G.K.}},
\bauthor{\bsnm{Panjabi}, \binits{R.}},
\bauthor{\bsnm{Nordstr{\"{o}}m}, \binits{A.}},
\bauthor{\bsnm{Legido-Quigley}, \binits{H.}}:
\batitle{{Health systems resilience in managing the COVID-19 pandemic: lessons from 28 countries}}.
\bjtitle{Nature Medicine}
\bvolume{27}(\bissue{6}),
\bfpage{964}--\blpage{980}
(\byear{2021}).
doi:\doiurl{10.1038/s41591-021-01381-y}
\end{barticle}
\endbibitem

\bibitem{Bertin2020}
\begin{barticle}
\bauthor{\bsnm{Bertin}, \binits{P.}},
\bauthor{\bsnm{Nera}, \binits{K.}},
\bauthor{\bsnm{Delouv{\'{e}}e}, \binits{S.}}:
\batitle{{Conspiracy Beliefs, Rejection of Vaccination, and Support for hydroxychloroquine: A Conceptual Replication-Extension in the COVID-19 Pandemic Context}}.
\bjtitle{Frontiers in Psychology}
\bvolume{11}(\bissue{September}),
\bfpage{1}--\blpage{9}
(\byear{2020}).
doi:\doiurl{10.3389/fpsyg.2020.565128}
\end{barticle}
\endbibitem

\end{thebibliography}

\newcommand{\BMCxmlcomment}[1]{}

\BMCxmlcomment{

<refgrp>

<bibl id="B1">
  <title><p>{Contested effects and chaotic policies: the 2020 story of (hydroxy) chloroquine for treating COVID-19}</p></title>
  <aug>
    <au><snm>Gould</snm><fnm>S</fnm></au>
    <au><snm>Norris</snm><fnm>SL</fnm></au>
  </aug>
  <source>The Cochrane database of systematic reviews</source>
  <pubdate>2021</pubdate>
  <volume>3</volume>
</bibl>

<bibl id="B2">
  <title><p>{Hydroxychloroquine for COVID19: The curtains close on a comedy of errors}</p></title>
  <aug>
    <au><snm>Schwartz</snm><fnm>IS</fnm></au>
    <au><snm>Boulware</snm><fnm>DR</fnm></au>
    <au><snm>Lee</snm><fnm>TC</fnm></au>
  </aug>
  <source>The Lancet Regional Health - Americas</source>
  <publisher>Elsevier Ltd</publisher>
  <pubdate>2022</pubdate>
  <volume>11</volume>
  <fpage>100268</fpage>
  <url>https://doi.org/10.1016/j.lana.2022.100268</url>
</bibl>

<bibl id="B3">
  <title><p>{Biomedicalization and the public sphere: Newspaper coverage ofhealth and medicine, 1960s-2000s}</p></title>
  <aug>
    <au><snm>Hallin</snm><fnm>DC</fnm></au>
    <au><snm>Brandt</snm><fnm>M</fnm></au>
    <au><snm>Briggs</snm><fnm>CL</fnm></au>
  </aug>
  <source>Social Science and Medicine</source>
  <publisher>Elsevier Ltd</publisher>
  <pubdate>2013</pubdate>
  <volume>96</volume>
  <fpage>121</fpage>
  <lpage>-128</lpage>
  <url>http://dx.doi.org/10.1016/j.socscimed.2013.07.030</url>
</bibl>

<bibl id="B4">
  <title><p>{Asymmetric participation of defenders and critics of vaccines to debates on French-speaking Twitter}</p></title>
  <aug>
    <au><snm>Gargiulo</snm><fnm>F</fnm></au>
    <au><snm>Cafiero</snm><fnm>F</fnm></au>
    <au><snm>Guille Escuret</snm><fnm>P</fnm></au>
    <au><snm>Seror</snm><fnm>V</fnm></au>
    <au><snm>Ward</snm><fnm>JK</fnm></au>
  </aug>
  <source>Scientific Reports</source>
  <pubdate>2020</pubdate>
  <volume>10</volume>
  <issue>1</issue>
  <fpage>6599</fpage>
  <url>http://arxiv.org/abs/1909.08311 https://www.nature.com/articles/s41598-020-62880-5</url>
</bibl>

<bibl id="B5">
  <title><p>{The Impact of the COVID-19 ``Infodemic'' on Drug-Utilization Behaviors: Implications for Pharmacovigilance}</p></title>
  <aug>
    <au><snm>Tuccori</snm><fnm>M</fnm></au>
    <au><snm>Convertino</snm><fnm>I</fnm></au>
    <au><snm>Ferraro</snm><fnm>S</fnm></au>
    <au><snm>Cappello</snm><fnm>E</fnm></au>
    <au><snm>Valdiserra</snm><fnm>G</fnm></au>
    <au><snm>Focosi</snm><fnm>D</fnm></au>
    <au><snm>Blandizzi</snm><fnm>C</fnm></au>
  </aug>
  <source>Drug Safety</source>
  <publisher>Springer International Publishing</publisher>
  <pubdate>2020</pubdate>
  <volume>43</volume>
  <issue>8</issue>
  <fpage>699</fpage>
  <lpage>-709</lpage>
  <url>https://doi.org/10.1007/s40264-020-00965-w</url>
</bibl>

<bibl id="B6">
  <title><p>{Hydroxychloroquine and COVID-19: Lack of Efficacy and the Social Construction of Plausibility}</p></title>
  <aug>
    <au><snm>Rughiniş</snm><fnm>C</fnm></au>
    <au><snm>Dima</snm><fnm>L</fnm></au>
    <au><snm>Vasile</snm><fnm>S</fnm></au>
  </aug>
  <source>American Journal of Therapeutics</source>
  <pubdate>2020</pubdate>
  <volume>27</volume>
  <issue>6</issue>
  <fpage>e573</fpage>
  <lpage>-e583</lpage>
  <url>https://journals.lww.com/10.1097/MJT.0000000000001294</url>
</bibl>

<bibl id="B7">
  <title><p>{A stance data set on polarized conversations on Twitter about the efficacy of hydroxychloroquine as a treatment for COVID-19}</p></title>
  <aug>
    <au><snm>Mutlu</snm><fnm>EC</fnm></au>
    <au><snm>Oghaz</snm><fnm>T</fnm></au>
    <au><snm>Jasser</snm><fnm>J</fnm></au>
    <au><snm>Tutunculer</snm><fnm>E</fnm></au>
    <au><snm>Rajabi</snm><fnm>A</fnm></au>
    <au><snm>Tayebi</snm><fnm>A</fnm></au>
    <au><snm>Ozmen</snm><fnm>O</fnm></au>
    <au><snm>Garibay</snm><fnm>I</fnm></au>
  </aug>
  <source>Data in Brief</source>
  <publisher>Elsevier Inc.</publisher>
  <pubdate>2020</pubdate>
  <volume>33</volume>
  <fpage>106401</fpage>
  <url>https://doi.org/10.1016/j.dib.2020.106401</url>
</bibl>

<bibl id="B8">
  <title><p>{How COVID broke the evidence pipeline}</p></title>
  <aug>
    <au><snm>Pearson</snm><fnm>H</fnm></au>
  </aug>
  <source>Nature</source>
  <pubdate>2021</pubdate>
  <volume>593</volume>
  <issue>7858</issue>
  <fpage>182</fpage>
  <lpage>-185</lpage>
  <url>http://www.nature.com/articles/d41586-021-01246-x</url>
</bibl>

<bibl id="B9">
  <title><p>{Science under Covid-19's magnifying glass: Lessons from the first months of the chloroquine debate in the French press}</p></title>
  <aug>
    <au><snm>Schultz</snm><fnm>{\'{E}}</fnm></au>
    <au><snm>Ward</snm><fnm>JK</fnm></au>
  </aug>
  <source>Journal of Sociology</source>
  <pubdate>2022</pubdate>
  <volume>58</volume>
  <issue>1</issue>
  <fpage>76</fpage>
  <lpage>-94</lpage>
  <url>http://journals.sagepub.com/doi/10.1177/1440783321999453</url>
</bibl>

<bibl id="B10">
  <title><p>{The Hydroxychloroquine Twitter War: A case study examining polarization in science communication}</p></title>
  <aug>
    <au><snm>Marcon</snm><fnm>AR</fnm></au>
    <au><snm>Caulfield</snm><fnm>T</fnm></au>
  </aug>
  <source>First Monday</source>
  <pubdate>2021</pubdate>
  <url>https://journals.uic.edu/ojs/index.php/fm/article/view/11707</url>
</bibl>

<bibl id="B11">
  <title><p>{Why Trust Raoult? How Social Indicators Inform the Reputations of Experts}</p></title>
  <aug>
    <au><snm>Origgi</snm><fnm>G</fnm></au>
    <au><snm>Branch smith</snm><fnm>T</fnm></au>
    <au><snm>Trust</snm><fnm>TM</fnm></au>
  </aug>
  <source>Social Epistemology</source>
  <pubdate>2022</pubdate>
</bibl>

<bibl id="B12">
  <title><p>{The ``Infodemic'' Infodemic: Toward a More Nuanced Understanding of Truth-Claims and the Need for (Not) Combatting Misinformation}</p></title>
  <aug>
    <au><snm>Krause</snm><fnm>NM</fnm></au>
    <au><snm>Freiling</snm><fnm>I</fnm></au>
    <au><snm>Scheufele</snm><fnm>DA</fnm></au>
  </aug>
  <source>Annals of the American Academy of Political and Social Science</source>
  <pubdate>2022</pubdate>
  <volume>700</volume>
  <issue>1</issue>
  <fpage>112</fpage>
  <lpage>-123</lpage>
</bibl>

<bibl id="B13">
  <title><p>{Death threats after a trial on chloroquine for COVID-19}</p></title>
  <aug>
    <au><snm>Ektorp</snm><fnm>E</fnm></au>
  </aug>
  <source>The Lancet. Infectious diseases</source>
  <publisher>Elsevier Ltd</publisher>
  <pubdate>2020</pubdate>
  <volume>20</volume>
  <issue>6</issue>
  <fpage>661</fpage>
  <url>http://dx.doi.org/10.1016/S1473-3099(20)30383-2</url>
</bibl>

<bibl id="B14">
  <title><p>{Hydroxychloroquine Controversies: Clinical Trials, Epistemology, and the Democratization of Science}</p></title>
  <aug>
    <au><snm>Berlivet</snm><fnm>L</fnm></au>
    <au><snm>L{\"{o}}wy</snm><fnm>I</fnm></au>
  </aug>
  <source>Medical Anthropology Quarterly</source>
  <pubdate>2020</pubdate>
  <volume>34</volume>
  <issue>4</issue>
  <fpage>525</fpage>
  <lpage>-541</lpage>
  <url>https://onlinelibrary.wiley.com/doi/10.1111/maq.12622</url>
</bibl>

<bibl id="B15">
  <title><p>{Does the public know when a scientific controversy is over? Public perceptions of hydroxychloroquine in France between April 2020 and June 2021}</p></title>
  <aug>
    <au><snm>Schultz</snm><fnm>{\'{E}}</fnm></au>
    <au><snm>Atlani Duault</snm><fnm>L</fnm></au>
    <au><snm>Peretti Watel</snm><fnm>P</fnm></au>
    <au><snm>Ward</snm><fnm>JK</fnm></au>
  </aug>
  <source>Therapies</source>
  <pubdate>2022</pubdate>
  <issue>January</issue>
  <url>https://linkinghub.elsevier.com/retrieve/pii/S0040595722000105</url>
</bibl>

<bibl id="B16">
  <title><p>{La controverse autour de Didier Raoult et de sa proposition th{\'{e}}rapeutique contre le Covid-19 sur Twitter : analyse de r{\'{e}}seaux et de discours}</p></title>
  <aug>
    <au><snm>Smyrnaios</snm><fnm>N</fnm></au>
    <au><snm>Tsimboukis</snm><fnm>P</fnm></au>
    <au><snm>Loub{\`{e}}re</snm><fnm>L</fnm></au>
  </aug>
  <source>Communiquer. Revue de communication sociale et publique</source>
  <pubdate>2021</pubdate>
  <volume>31</volume>
  <fpage>1</fpage>
  <lpage>-20</lpage>
</bibl>

<bibl id="B17">
  <title><p>Le contrôle par les pairs au temps du coronavirus</p></title>
  <aug>
    <au><snm>Dubois</snm><fnm>M</fnm></au>
    <au><snm>Frenod Dunaud</snm><fnm>A</fnm></au>
    <au><snm>Gargiulo</snm><fnm>F</fnm></au>
    <au><snm>Guaspare Carton</snm><fnm>C</fnm></au>
  </aug>
  <source>Sorbonnavirus, Regards sur la crise du coronavirus</source>
  <publisher>Sorbonne Université Presses</publisher>
  <editor>Chauvin, P.M. and Clement, A</editor>
  <pubdate>2021</pubdate>
</bibl>

<bibl id="B18">
  <title><p>{Political Prescriptions: Three Pandemic Stories}</p></title>
  <aug>
    <au><snm>Bharti</snm><fnm>N</fnm></au>
    <au><snm>Sismondo</snm><fnm>S</fnm></au>
  </aug>
  <source>Science, Technology \& Human Values</source>
  <pubdate>2022</pubdate>
  <fpage>016224392211238</fpage>
  <url>http://journals.sagepub.com/doi/10.1177/01622439221123831</url>
</bibl>

<bibl id="B19">
  <title><p>{Covid-19: En Afrique de l'Ouest, le vaccin n'est pas le nouveau ``magic bullet''}</p></title>
  <aug>
    <au><snm>Desclaux</snm><fnm>A</fnm></au>
  </aug>
  <pubdate>2022</pubdate>
  <url>https://vih.org/20210202/la-mondialisation-des-informations-et-la-fabrique-des-opinions-sur-les-traitements-du-covid-en-afrique/</url>
</bibl>

<bibl id="B20">
  <title><p>{Mortality outcomes with hydroxychloroquine and chloroquine in COVID-19 from an international collaborative meta-analysis of randomized trials}</p></title>
  <aug>
    <au><snm>Axfors</snm><fnm>C</fnm></au>
    <au><snm>Schmitt</snm><fnm>AM</fnm></au>
    <au><snm>Janiaud</snm><fnm>P</fnm></au>
    <au><snm>Hooft</snm><fnm>J</fnm></au>
    <au><snm>Abd Elsalam</snm><fnm>S</fnm></au>
    <au><snm>Abdo</snm><fnm>EF</fnm></au>
    <au><snm>Abella</snm><fnm>BS</fnm></au>
    <au><snm>Akram</snm><fnm>J</fnm></au>
    <au><snm>Amaravadi</snm><fnm>RK</fnm></au>
    <au><snm>Angus</snm><fnm>DC</fnm></au>
    <au><snm>Arabi</snm><fnm>YM</fnm></au>
    <au><snm>Azhar</snm><fnm>S</fnm></au>
    <au><snm>Baden</snm><fnm>LR</fnm></au>
    <au><snm>Baker</snm><fnm>AW</fnm></au>
    <au><snm>Belkhir</snm><fnm>L</fnm></au>
    <au><snm>Benfield</snm><fnm>T</fnm></au>
    <au><snm>Berrevoets</snm><fnm>MA</fnm></au>
    <au><snm>Chen</snm><fnm>CP</fnm></au>
    <au><snm>Chen</snm><fnm>TC</fnm></au>
    <au><snm>Cheng</snm><fnm>SH</fnm></au>
    <au><snm>Cheng</snm><fnm>CY</fnm></au>
    <au><snm>Chung</snm><fnm>WS</fnm></au>
    <au><snm>Cohen</snm><fnm>YZ</fnm></au>
    <au><snm>Cowan</snm><fnm>LN</fnm></au>
    <au><snm>Dalgard</snm><fnm>O</fnm></au>
    <au><snm>{de Almeida e Val}</snm><fnm>FF</fnm></au>
    <au><snm>Lacerda</snm><fnm>MV</fnm></au>
    <au><snm>Melo</snm><fnm>GC</fnm></au>
    <au><snm>Derde</snm><fnm>L</fnm></au>
    <au><snm>Dubee</snm><fnm>V</fnm></au>
    <au><snm>Elfakir</snm><fnm>A</fnm></au>
    <au><snm>Gordon</snm><fnm>AC</fnm></au>
    <au><snm>Hernandez Cardenas</snm><fnm>CM</fnm></au>
    <au><snm>Hills</snm><fnm>T</fnm></au>
    <au><snm>Hoepelman</snm><fnm>AI</fnm></au>
    <au><snm>Huang</snm><fnm>YW</fnm></au>
    <au><snm>Igau</snm><fnm>B</fnm></au>
    <au><snm>Jin</snm><fnm>R</fnm></au>
    <au><snm>Jurado Camacho</snm><fnm>F</fnm></au>
    <au><snm>Khan</snm><fnm>KS</fnm></au>
    <au><snm>Kremsner</snm><fnm>PG</fnm></au>
    <au><snm>Kreuels</snm><fnm>B</fnm></au>
    <au><snm>Kuo</snm><fnm>CY</fnm></au>
    <au><snm>Le</snm><fnm>T</fnm></au>
    <au><snm>Lin</snm><fnm>YC</fnm></au>
    <au><snm>Lin</snm><fnm>WP</fnm></au>
    <au><snm>Lin</snm><fnm>TH</fnm></au>
    <au><snm>Lyngbakken</snm><fnm>MN</fnm></au>
    <au><snm>McArthur</snm><fnm>C</fnm></au>
    <au><snm>McVerry</snm><fnm>BJ</fnm></au>
    <au><snm>Meza Meneses</snm><fnm>P</fnm></au>
    <au><snm>Monteiro</snm><fnm>WM</fnm></au>
    <au><snm>Morpeth</snm><fnm>SC</fnm></au>
    <au><snm>Mourad</snm><fnm>A</fnm></au>
    <au><snm>Mulligan</snm><fnm>MJ</fnm></au>
    <au><snm>Murthy</snm><fnm>S</fnm></au>
    <au><snm>Naggie</snm><fnm>S</fnm></au>
    <au><snm>Narayanasamy</snm><fnm>S</fnm></au>
    <au><snm>Nichol</snm><fnm>A</fnm></au>
    <au><snm>Novack</snm><fnm>LA</fnm></au>
    <au><snm>O'Brien</snm><fnm>SM</fnm></au>
    <au><snm>Okeke</snm><fnm>NL</fnm></au>
    <au><snm>Perez</snm><fnm>L</fnm></au>
    <au><snm>Perez Padilla</snm><fnm>R</fnm></au>
    <au><snm>Perrin</snm><fnm>L</fnm></au>
    <au><snm>Remigio Luna</snm><fnm>A</fnm></au>
    <au><snm>Rivera Martinez</snm><fnm>NE</fnm></au>
    <au><snm>Rockhold</snm><fnm>FW</fnm></au>
    <au><snm>Rodriguez Llamazares</snm><fnm>S</fnm></au>
    <au><snm>Rolfe</snm><fnm>R</fnm></au>
    <au><snm>Rosa</snm><fnm>R</fnm></au>
    <au><snm>R{\o}sj{\o}</snm><fnm>H</fnm></au>
    <au><snm>Sampaio</snm><fnm>VS</fnm></au>
    <au><snm>Seto</snm><fnm>TB</fnm></au>
    <au><snm>Shehzad</snm><fnm>M</fnm></au>
    <au><snm>Soliman</snm><fnm>S</fnm></au>
    <au><snm>Stout</snm><fnm>JE</fnm></au>
    <au><snm>Thirion Romero</snm><fnm>I</fnm></au>
    <au><snm>Troxel</snm><fnm>AB</fnm></au>
    <au><snm>Tseng</snm><fnm>TY</fnm></au>
    <au><snm>Turner</snm><fnm>NA</fnm></au>
    <au><snm>Ulrich</snm><fnm>RJ</fnm></au>
    <au><snm>Walsh</snm><fnm>SR</fnm></au>
    <au><snm>Webb</snm><fnm>SA</fnm></au>
    <au><snm>Weehuizen</snm><fnm>JM</fnm></au>
    <au><snm>Velinova</snm><fnm>M</fnm></au>
    <au><snm>Wong</snm><fnm>HL</fnm></au>
    <au><snm>Wrenn</snm><fnm>R</fnm></au>
    <au><snm>Zampieri</snm><fnm>FG</fnm></au>
    <au><snm>Zhong</snm><fnm>W</fnm></au>
    <au><snm>Moher</snm><fnm>D</fnm></au>
    <au><snm>Goodman</snm><fnm>SN</fnm></au>
    <au><snm>Ioannidis</snm><fnm>JP</fnm></au>
    <au><snm>Hemkens</snm><fnm>LG</fnm></au>
  </aug>
  <source>Nature Communications</source>
  <pubdate>2021</pubdate>
  <volume>12</volume>
  <issue>1</issue>
  <fpage>1</fpage>
  <lpage>-13</lpage>
</bibl>

<bibl id="B21">
  <title><p>{From hydroxychloroquine to ivermectin: how unproven ``cures'' can go viral}</p></title>
  <aug>
    <au><snm>Taccone</snm><fnm>FS</fnm></au>
    <au><snm>Hites</snm><fnm>M</fnm></au>
    <au><snm>Dauby</snm><fnm>N</fnm></au>
  </aug>
  <source>Clinical Microbiology and Infection</source>
  <publisher>European Society of Clinical Microbiology and Infectious Diseases</publisher>
  <pubdate>2022</pubdate>
  <volume>28</volume>
  <issue>4</issue>
  <fpage>472</fpage>
  <lpage>-474</lpage>
  <url>https://doi.org/10.1016/j.cmi.2022.01.008</url>
</bibl>

<bibl id="B22">
  <title><p>{The hydroxychloroquine alliance: how far-right leaders and alt-science preachers came together to promote a miracle drug}</p></title>
  <aug>
    <au><snm>Casar{\~{o}}es</snm><fnm>G</fnm></au>
    <au><snm>Magalh{\~{a}}es</snm><fnm>D</fnm></au>
  </aug>
  <source>Revista de Administracao Publica</source>
  <pubdate>2021</pubdate>
  <volume>55</volume>
  <issue>1</issue>
  <fpage>197</fpage>
  <lpage>-214</lpage>
</bibl>

<bibl id="B23">
  <title><p>{Opinion and uptake of chloroquine for treatment of COVID-19 during the mandatory lockdown in the sub-Saharan African region}</p></title>
  <aug>
    <au><snm>Osuagwu</snm><fnm>UL</fnm></au>
    <au><snm>Nwaeze</snm><fnm>O</fnm></au>
    <au><snm>Ovenseri Ogbomo</snm><fnm>G</fnm></au>
    <au><snm>Oloruntoba</snm><fnm>R</fnm></au>
    <au><snm>Ekpenyong</snm><fnm>B</fnm></au>
    <au><snm>Mashige</snm><fnm>KP</fnm></au>
    <au><snm>Timothy</snm><fnm>C</fnm></au>
    <au><snm>Ishaya</snm><fnm>T</fnm></au>
    <au><snm>Langsi</snm><fnm>R</fnm></au>
    <au><snm>Charwe</snm><fnm>D</fnm></au>
    <au><snm>Abu</snm><fnm>EK</fnm></au>
    <au><snm>Chundung</snm><fnm>MA</fnm></au>
    <au><snm>Agho</snm><fnm>KE</fnm></au>
  </aug>
  <source>African Journal of Primary Health Care \& Family Medicine</source>
  <pubdate>2021</pubdate>
  <volume>13</volume>
  <issue>1</issue>
  <fpage>1</fpage>
  <lpage>-8</lpage>
  <url>http://www.phcfm.org/index.php/PHCFM/article/view/2795</url>
</bibl>

<bibl id="B24">
  <title><p>{COVID-19 Treatment Protocols in the WHO African Region - Results of a Survey}</p></title>
  <aug>
    <au><snm>Mukankubito</snm><fnm>I</fnm></au>
    <au><snm>Annan</snm><fnm>EA</fnm></au>
    <au><snm>Sougou</snm><fnm>A</fnm></au>
    <au><snm>Taguembou</snm><fnm>D</fnm></au>
    <au><snm>Kniazkov</snm><fnm>S</fnm></au>
    <au><snm>Loua</snm><fnm>A</fnm></au>
    <au><snm>Mankele</snm><fnm>R</fnm></au>
    <au><snm>Nikiema</snm><fnm>JB</fnm></au>
    <au><snm>Bisoborwa</snm><fnm>G</fnm></au>
    <au><snm>Julius</snm><fnm>OKM</fnm></au>
  </aug>
  <pubdate>2021</pubdate>
  <fpage>1</fpage>
  <lpage>-19</lpage>
  <url>http://europepmc.org/abstract/PPR/PPR339906
</bibl>

<bibl id="B25">
  <title><p>{State humanitarian verticalism versus universal health coverage: A century of French international health assistance revisited}</p></title>
  <aug>
    <au><snm>Atlani Duault</snm><fnm>L</fnm></au>
    <au><snm>Dozon</snm><fnm>JP</fnm></au>
    <au><snm>Wilson</snm><fnm>A</fnm></au>
    <au><snm>Delfraissy</snm><fnm>JF</fnm></au>
    <au><snm>Moatti</snm><fnm>JP</fnm></au>
  </aug>
  <source>The Lancet</source>
  <pubdate>2016</pubdate>
  <volume>387</volume>
  <issue>10034</issue>
  <fpage>2250</fpage>
  <lpage>-2262</lpage>
</bibl>

<bibl id="B26">
  <title><p>{COVID-19 Epidemic: Chloroquine, a French Obsession?}</p></title>
  <aug>
    <au><snm>Lapostolle</snm><fnm>F</fnm></au>
    <au><snm>Vianu</snm><fnm>I</fnm></au>
    <au><snm>{De Stefano}</snm><fnm>C</fnm></au>
    <au><snm>Goix</snm><fnm>L</fnm></au>
    <au><snm>Petrovic</snm><fnm>T</fnm></au>
    <au><snm>Adnet</snm><fnm>F</fnm></au>
  </aug>
  <source>La Presse M{\'{e}}dicale Open</source>
  <publisher>Elsevier Masson SAS</publisher>
  <pubdate>2021</pubdate>
  <volume>2</volume>
  <fpage>100007</fpage>
  <url>https://doi.org/10.1016/j.lpmope.2021.100007</url>
</bibl>

<bibl id="B27">
  <title><p>{Ebola and Localized Blame on Social Media: Analysis of Twitter and Facebook Conversations During the 2014--2015 Ebola Epidemic}</p></title>
  <aug>
    <au><snm>Roy</snm><fnm>M</fnm></au>
    <au><snm>Moreau</snm><fnm>N</fnm></au>
    <au><snm>Rousseau</snm><fnm>C</fnm></au>
    <au><snm>Mercier</snm><fnm>A</fnm></au>
    <au><snm>Wilson</snm><fnm>A</fnm></au>
    <au><snm>Atlani Duault</snm><fnm>L</fnm></au>
  </aug>
  <source>Culture, Medicine and Psychiatry</source>
  <pubdate>2020</pubdate>
  <volume>44</volume>
  <issue>1</issue>
  <fpage>56</fpage>
  <lpage>-79</lpage>
</bibl>

<bibl id="B28">
  <title><p>Gazouilloire, Twitter stream + search API grabber</p></title>
  <aug>
    <au><snm>Ooghe Tabanou</snm><fnm>B</fnm></au>
    <au><snm>Farjas</snm><fnm>J</fnm></au>
    <au><snm>Mazoyer</snm><fnm>B</fnm></au>
  </aug>
  <url>https://github.com/medialab/gazouilloire</url>
</bibl>

<bibl id="B29">
  <title><p>maurofaccin/DataCovVac</p></title>
  <aug>
    <au><snm>Faccin</snm><fnm>M</fnm></au>
  </aug>
  <publisher>Zenodo</publisher>
  <pubdate>2023</pubdate>
  <url>https://doi.org/10.5281/zenodo.7870249</url>
</bibl>

<bibl id="B30">
  <title><p>Assessing the influence of French vaccine critics during the two first years of the { COVID}-19 pandemic</p></title>
  <aug>
    <au><snm>Faccin</snm><fnm>M</fnm></au>
    <au><snm>Gargiulo</snm><fnm>F</fnm></au>
    <au><snm>Atlani Duault</snm><fnm>L</fnm></au>
    <au><snm>Ward</snm><fnm>JK</fnm></au>
  </aug>
  <source>arxiv:2202.10952</source>
  <pubdate>2022</pubdate>
  <url>http://arxiv.org/abs/2202.10952v1</url>
</bibl>

<bibl id="B31">
  <title><p>maurofaccin/DataCovHCQ</p></title>
  <aug>
    <au><snm>Faccin</snm><fnm>M</fnm></au>
  </aug>
  <publisher>Zenodo</publisher>
  <pubdate>2023</pubdate>
  <url>https://doi.org/10.5281/zenodo.7870286</url>
</bibl>

<bibl id="B32">
  <title><p>Directed hypergraphs: Introduction and fundamental algorithms{\textemdash}A survey</p></title>
  <aug>
    <au><snm>Ausiello</snm><fnm>G</fnm></au>
    <au><snm>Laura</snm><fnm>L</fnm></au>
  </aug>
  <source>Theoretical Computer Science</source>
  <publisher>Elsevier {BV}</publisher>
  <pubdate>2017</pubdate>
  <volume>658</volume>
  <fpage>293</fpage>
  <lpage>-306</lpage>
  <url>https://doi.org/10.1016
</bibl>

<bibl id="B33">
  <title><p>Measuring dynamical systems on directed hypergraphs</p></title>
  <aug>
    <au><snm>Faccin</snm><fnm>M</fnm></au>
  </aug>
  <source>arxiv:2202.12810</source>
  <pubdate>2022</pubdate>
  <url>http://arxiv.org/abs/2202.12810v1</url>
</bibl>

<bibl id="B34">
  <title><p>Non-negative tensor factorization with applications to statistics and computer vision</p></title>
  <aug>
    <au><snm>Shashua</snm><fnm>A</fnm></au>
    <au><snm>Hazan</snm><fnm>T</fnm></au>
  </aug>
  <source>Proceedings of the 22nd international conference on Machine learning - {ICML} 2005</source>
  <publisher>{ACM} Press</publisher>
  <pubdate>2005</pubdate>
  <url>https://doi.org/10.1145/1102351.1102451</url>
</bibl>

<bibl id="B35">
  <title><p>Nonnegative Matrix and Tensor Factorizations</p></title>
  <aug>
    <au><snm>Cichocki</snm><fnm>A</fnm></au>
    <au><snm>Zdunek</snm><fnm>R</fnm></au>
    <au><snm>Phan</snm><fnm>AH</fnm></au>
    <au><snm>Amari</snm><fnm>SI</fnm></au>
  </aug>
  <publisher>John Wiley {\&} Sons, Ltd</publisher>
  <pubdate>2009</pubdate>
  <url>https://doi.org/10.1002/9780470747278</url>
</bibl>

<bibl id="B36">
  <title><p>Applications of tensor (multiway array) factorizations and decompositions in data mining</p></title>
  <aug>
    <au><snm>M{\o}rup</snm><fnm>M</fnm></au>
  </aug>
  <source>WIREs Data Mining and Knowledge Discovery</source>
  <publisher>Wiley</publisher>
  <pubdate>2011</pubdate>
  <volume>1</volume>
  <issue>1</issue>
  <fpage>24</fpage>
  <lpage>-40</lpage>
  <url>https://doi.org/10.1002/widm.1</url>
</bibl>

<bibl id="B37">
  <title><p>Detecting the Community Structure and Activity Patterns of Temporal Networks: A Non-Negative Tensor Factorization Approach</p></title>
  <aug>
    <au><snm>Gauvin</snm><fnm>L</fnm></au>
    <au><snm>Panisson</snm><fnm>A</fnm></au>
    <au><snm>Cattuto</snm><fnm>C</fnm></au>
  </aug>
  <source>PLoS {ONE}</source>
  <publisher>Public Library of Science ({PLoS})</publisher>
  <editor>Yamir Moreno</editor>
  <pubdate>2014</pubdate>
  <volume>9</volume>
  <issue>1</issue>
  <fpage>e86028</fpage>
  <url>https://doi.org/10.1371
</bibl>

<bibl id="B38">
  <title><p>{Health systems resilience in managing the COVID-19 pandemic: lessons from 28 countries}</p></title>
  <aug>
    <au><snm>Haldane</snm><fnm>V</fnm></au>
    <au><snm>{De Foo}</snm><fnm>C</fnm></au>
    <au><snm>Abdalla</snm><fnm>SM</fnm></au>
    <au><snm>Jung</snm><fnm>AS</fnm></au>
    <au><snm>Tan</snm><fnm>M</fnm></au>
    <au><snm>Wu</snm><fnm>S</fnm></au>
    <au><snm>Chua</snm><fnm>A</fnm></au>
    <au><snm>Verma</snm><fnm>M</fnm></au>
    <au><snm>Shrestha</snm><fnm>P</fnm></au>
    <au><snm>Singh</snm><fnm>S</fnm></au>
    <au><snm>Perez</snm><fnm>T</fnm></au>
    <au><snm>Tan</snm><fnm>SM</fnm></au>
    <au><snm>Bartos</snm><fnm>M</fnm></au>
    <au><snm>Mabuchi</snm><fnm>S</fnm></au>
    <au><snm>Bonk</snm><fnm>M</fnm></au>
    <au><snm>McNab</snm><fnm>C</fnm></au>
    <au><snm>Werner</snm><fnm>GK</fnm></au>
    <au><snm>Panjabi</snm><fnm>R</fnm></au>
    <au><snm>Nordstr{\"{o}}m</snm><fnm>A</fnm></au>
    <au><snm>Legido Quigley</snm><fnm>H</fnm></au>
  </aug>
  <source>Nature Medicine</source>
  <publisher>Springer US</publisher>
  <pubdate>2021</pubdate>
  <volume>27</volume>
  <issue>6</issue>
  <fpage>964</fpage>
  <lpage>-980</lpage>
  <url>http://dx.doi.org/10.1038/s41591-021-01381-y</url>
</bibl>

<bibl id="B39">
  <title><p>{Conspiracy Beliefs, Rejection of Vaccination, and Support for hydroxychloroquine: A Conceptual Replication-Extension in the COVID-19 Pandemic Context}</p></title>
  <aug>
    <au><snm>Bertin</snm><fnm>P</fnm></au>
    <au><snm>Nera</snm><fnm>K</fnm></au>
    <au><snm>Delouv{\'{e}}e</snm><fnm>S</fnm></au>
  </aug>
  <source>Frontiers in Psychology</source>
  <pubdate>2020</pubdate>
  <volume>11</volume>
  <issue>September</issue>
  <fpage>1</fpage>
  <lpage>-9</lpage>
</bibl>

</refgrp>
} 


\end{backmatter}
\end{document}